# Skills or Degree? The Rise of Skill-Based Hiring for AI and Green Jobs*

Matthew Bone◇     Eugenia Ehlinger♣     Fabian Stephany♠✉

November 2024

## Abstract

Emerging professions in fields like Artificial Intelligence (AI) and sustainability (green jobs) are experiencing labour shortages as industry demand outpaces labour supply. In this context, our study aims to understand whether employers have begun focusing more on individual skills rather than formal qualifications in their recruitment processes. We analysed a large time-series dataset of approximately eleven million online job vacancies in the UK from 2018 to mid-2024, drawing on diverse literature on technological change and labour market signalling. Our findings provide evidence that employers have initiated "skill-based hiring" for AI roles, adopting more flexible hiring practices to expand the available talent pool. From 2018-2023, demand for AI roles grew by 21% as a proportion of all postings (and accelerated into 2024). Simultaneously, mentions of university education requirements for AI roles declined by 15%. Our regression analysis shows that university degrees have a significantly lower wage premium for both AI and green roles. In contrast, AI skills command a wage premium of 23%, exceeding the value of degrees up until the PhD-level (33%). In occupations with high demand for AI skills, the premium for skills is high, and the reward for degrees is relatively low. We recommend leveraging alternative skill-building formats such as apprenticeships, on-the-job training, MOOCs, vocational education and training, micro-certificates, and online bootcamps to fully utilise human capital and address talent shortages.

*Keywords*: Future of Work, Labour Markets, Skills, Education, AI, Sustainability
*JEL Class*: C55, I23 J23, J24, J31

♣ UNESCO, Paris,
◇ Oxford Internet Institute & Burning Glass Institute
♠ Oxford Internet Institute, University of Oxford, UK; Humboldt Institute for Internet and Society, Berlin; Bruegel, Brussels, Belgium; DWG Berlin; ✉ fabian.stephany@oii.ox.ac.uk
*This work is part of the Buegel Future of Work and Inclusive Growth project. The authors are very grateful to the feedback received on the project's 2023 seminar and a friendly review by Estrella Gómez Herrera and Duygu Güner.

**Introduction**

Are skills more relevant than formal education? While educational qualifications often play an important role from an employer's perspective when hiring new employees, recent studies suggest that increasing attention is being paid to candidates' skills acquired outside formal education (Fuller, 2022; McKinsey, 2022; LinkedIn, 2022). This shift towards skills-focused hiring arises from the mismatch between the pace of evolution in formal education and the labour market, as well as the challenges companies face in finding and retaining suitable candidates when they limit themselves to individuals with higher educational degrees in specific fields. Observational studies have also indicated that employers are more likely to adopt hiring strategies focusing on individual skills when the demand for talent far exceeds supply.

For example, this practice was evident during the unprecedented need for medical staff amid the COVID-19 pandemic (Fuller, 2022). In response to the immense healthcare challenges posed by the pandemic, many employers demonstrated a willingness to temporarily suspend diploma prerequisites for various roles. For instance, the proportion of job listings for ICU and critical-care nurses requiring a bachelor's degree decreased by 12 percentage points from 2019 to 2020, falling from 35% to 23% (Fuller, 2022). While this change may be a temporary measure addressing urgent needs—viewed as a cyclical rather than structural adjustment—it is extensive enough to offer significant insights into whether degree holders actually outperform newly recruited counterparts without degrees. Similar to the short-term labour market shocks during the COVID-19 pandemic that prompted a rise in skill-based hiring practices, long-term trends are also indicating the need to broaden the talent pool through similar strategies.

Two of the most prominent long-term trends shaping today's labour market are the transition towards environmentally sustainable jobs and the technological revolution driven by the advent of AI. The so-called green transition—increased adoption of environmentally friendly practices and production—is rapidly reshaping industries, leading to a higher demand for jobs that support a more sustainable economy. This shift involves not only creating new jobs but also transforming existing ones, as highlighted by projections like those from the IEA, which foresee a 3.5% boost in global GDP and the creation of millions of jobs (IEA, 2020). Similarly, the technological wave, particularly the integration of AI into the workforce, is generating entirely new roles and fundamentally altering the skill sets required for employment. The advent of AI is reshaping the labour market, demanding new skill sets while leading to the paradox of job vacancies amid unemployment (Autor, 2015; Acemoglu et al., 2022). The rapid pace of these shifts necessitates new approaches to skill development and acquisition, as the conventional alignment of training with labour market demands struggles to remain relevant (Collins and Halverson, 2018; Illanes et al., 2018).

Both short-term shocks, such as the COVID-19 pandemic, and long-term transitions, like the shift to green jobs or technological change, are believed to drive skill-based hiring as firms struggle to access desperately needed talent with the corresponding formal education. Despite the growing importance of skills-based hiring and its potential to alleviate skills shortages by enlarging the talent pool, our understanding of it remains limited. Most prior scientific articles on employers' requirements focus on understanding changing trends in the skills demanded, without comparing the value placed on skills versus formal education. While these approaches are useful for providing insights into the evolving demands of the labour market and identifying emerging occupations and skills (Alekseva, 2021; Saussay, 2021; ILO, 2021), they often overlook the interplay between formal education and skills in the hiring process, thereby simplifying the complex dynamics among formal degrees, work experience, and skills from the employers' perspective.

We advance this discussion by analysing a large time-series dataset comprising about eleven million online job vacancies in the UK between January 2018 and June 2024, which includes granular details for every posting. We define skill-based hiring as "...the practice of employers setting specific skill or competency requirements or targets (for hiring)" (Butrica and Mudrazija, 2022). Focusing on AI-related and green jobs allows us to explore a simple premise: when employers require novel skills for which they have reported shortages, they adopt skills-based hiring to enlarge the talent pool. Accordingly, we pose the research question: "**Given the fast-growing demand, have employers started to apply skill-based hiring practices for AI and green jobs?**" We investigate whether firms are trying to tap into the talent pool by placing stronger emphasis and remuneration on individual in-demand skills rather than on formal education requirements. Using statistical analyses and a regression model, we examine the association between higher education degrees and skills requirements with economic remuneration. This enables us to provide new insights into the paradigm shift from a degree-focused to a skills-based labour market.

Based on our unique dataset on employers' requirements in job advertisements, we find evidence that jobs requiring AI and green skills are increasing and generally demand a broader set of skills. Our results show that vacancies demanding these disruptive skills are also associated with significant wage premiums. Furthermore, our findings suggest that for AI roles, skill requirements surpass the wage impact of educational attainment up to the PhD level. At the same time, for both AI and green roles, we observe that the premium for formal degrees has diminished. For the case of AI roles, we reveal that premia on skills are most pronounced for occupations with above-average demand for AI talent.

The paper proceeds as follows. Section 2 provides background literature and a conceptual framework that explicitly links human capital, in-demand skills, and wages. Section 3 describes our dataset and outlines the bottom-up approach we use to identify AI and green jobs, as well as the changing demand and supply in these two domains that might lead to a shift in hiring practices. Section 4 examines the aspect of human capital—central to our investigation—which is represented by formal education and skills. Here, we analyse and compare the educational and skills requirements and estimate their influence on wage premiums. Section 5 concludes with policy recommendations.

## Background

Our framework for examining whether employers have started focusing on individual skills rather than formal qualifications in recruiting for emerging occupations draws on human capital theory and the emerging economics of skills. The key forces in our conceptual framework are the excess demand for specific skills due to societal and technological transformations, and the resulting skill shortages faced by employers. Our investigation also incorporates concepts such as educational premium and skills-based hiring. In this way, we combine long-standing theories with novel research methodologies to better understand emerging practices in the labour market.

### *The emerging economics of skills*

Contemporary studies on the economic value of skills in the labour market often build upon the theory of human capital. The premise of this theory is straightforward: income results from human capital, meaning that knowledge and skills enhance productivity (Becker, 1994; Schultz, 1972; Holden & Biddle, 2017). Human capital theory has been applied across various strands of research, leading to diverse complementary or alternative perspectives. Since its inception, researchers have represented human capital in ways beyond or in addition to formal education. For example, Schultz (1972) views human capital as the capacity to adapt to changing circumstances and suggests that it is valuable primarily because it increases firms' productivity. Alternatively, Becker (1994) conceptualises skills as a durable investment acquired through schooling and on-the-job training. This viewpoint has enabled researchers to study the price of skills using experience and educational attainment as measures of investment, facilitating an understanding of the returns on education (see Mincer, 1974; Goldin & Katz, 2008).

Increasingly, representing human capital solely through education is being contested, as formal educational institutions struggle to keep pace with rapidly changing labour market requirements and technological advancements (Lutz et al., 2021; Aleksava, 2021). Interpreting skills, rather than formal education, as an economic production factor within the extended theory of human capital leads to various subsequent economic considerations. Is there a market for skills? Is this market governed by supply and demand? Do skills have a price?

Several recent studies have utilised more granular, skill-related data to better understand the economics of skills and how changes in supply and demand affect the price of a skill. For example, Stephany and Teutloff (2024) show how the premium on skills—that is, the additional contribution to a worker's wage—reacts to changes in supply and demand. In their model, using data from 25,000 knowledge workers and 962 related skills, a one per cent increase in supply decreases the wage premium of a skill by 27 percentage points, while a one per cent increase in demand elevates the value of a skill by 20 percentage points. Conversely, examining online platform workers, Duch-Brown et al. (2021) illustrate that demand and supply for specific skills exhibit particular levels of elasticity when the market faces an exogenous shock of an increase in the price of a skill.

Other scholars have studied how existing skill supply impacts employers' requirements (Goldin & Katz, 2008; Beaudry et al., 2013; Fuller & Raman, 2017; Burning Glass Institute, 2022). Modestino et al. (2020) demonstrate that when workforce supply exceeds firms' demand, employers require more skills and higher levels of education—a phenomenon evident in the US during the 2008 economic crisis, where workforce supply outpaced employers' demand due to macroeconomic conditions. Conversely, when workers are scarce and demand exceeds supply, we can expect the opposite effect: employers dropping degree requirements to enlarge the talent pool, as suggested by research investigating the impact of the COVID-19 pandemic (Fuller et al., 2022; Burning Glass Institute, 2022).

### The rise of green and AI roles

Just as the abrupt disruptions to the labour market during the COVID pandemic led to an increase in skill-focused recruitment, enduring trends are indicating a similar expansion of the talent pool is necessary, as illustrated with two major long-term movements: the shift toward jobs that prioritise environmental sustainability, and the technological overhaul brought about by the emergence of AI. For both trends the challenge is that the skill specifications for emerging jobs are often uncertain and in flux. Given this uncertainty and the rapid pace of technological and societal changes, traditional approaches to address skill mismatches—like aligning training with market needs—are losing effectiveness, as they lag behind the transformations (Collins and Halverson, 2018). Large corporations also face difficulties in maintaining a workforce whose skills are current (Illanes et al., 2018). We continue this stream of literature by challenging models that uniquely use schooling and formal education as proxies to human capital and instead use individual skills as a countermeasure.

The transition to a green economy, forecasted to stimulate both labour disruption and job creation, underscores the growing appeal of skill-based hiring. The IEA estimates that a green recovery could boost global GDP by 3.5% and lead to the creation of 9 million jobs each year (IEA, 2020). This shift is expected to raise employment by 1.2% by 2030 as a result of the European Green Deal (CEDEFOP, 2021) and has increased the demand for green skills, in addition to educational requirements, especially in high-skilled sectors (Vona et al., 2018; Marin and Vona, 2019). With the decline in traditional roles and their associated educational curricular, particularly in industries like mining and fossil fuels, and the burgeoning need for technological expertise in STEM fields, skill-based hiring becomes increasingly attractive as a means to bridge the skill gap and meet the dynamic demands of the evolving labour market (Buchanan, Kronk, 2023; Quintini and Venn, 2013).

Similar to a "greening" of the economy, technology impacts the labour market in a way that is far from "skill neutral"; it reshapes and creates jobs, each with unique skill requirements beyond formal education (Acemoglu and Autor, 2011; Brynjolfsson and Mitchell, 2017; Frey and Osborne, 2017). The adoption of AI has entailed conflicting views between academics. Some researchers suggest that AI will complement human capacities– from this point of view– we could assume that AI will drive an increase in the demand for workers. On the other hand, some scholars suggest that AI has the potential to replace humans, by completing tasks autonomously, which could, as a result, decrease the need for human workers (Acemoglu et al, 2022).

Moreover, AI, at the moment, seems to be fostering the generation of roles with entirely new tasks that demand innovative skill sets. This evolution leads to the ironic dilemma of experiencing job vacancies and unemployment simultaneously (Autor, 2015) – businesses can't find the right talent for novel positions as workers are displaced from traditional roles. Workers must now acquire new competencies and blend them creatively with their existing ones, while employers should focus on retraining their staff and seeking new talent.

### Is the workforce ready for change?

While our current study focuses on the demand side, the literature consistently underscores that the workforce is not meeting the needs of the economy and that the demand for talent is far outreaching skills supply, particularly in novel domains. Francis-Devine & Buchanan (2023) highlight that employers reported prevalent skills shortages, meaning people do not have the capacities needed in the labour market. These skills shortages present several economic and societal challenges. Further economic studies have shown that in the absence of skilled workers, businesses struggle to implement and develop new technologies, innovate and compete, which hinders economic development and exacerbates social inequalities (Strietska, 2007). Moreover, skills shortages polarise the labour market by limiting the number of high-paying jobs and restraining social mobility (UK Commission for Employment and Skills, 2014).

The UK is already facing imbalances in the labour market, with skills shortages posing significant difficulties. Research shows that the UK has a high percentage of STEM graduates compared to other countries with 42% of graduates completing STEM programmes. Yet, despite this, employers report difficulties in finding individuals with digital skills, which are needed to foster the development of advanced technologies such as AI (Cambridge Industrial Innovation Policy, 2024). For example, while the government has focused on fostering AI and technology startups, leading to cities like Cambridge and London becoming AI global hubs, employers in these fields report a lack of proficient workers, in particular for advanced digital skills. Beyond that, the scarcity of digital skills has become a prevalent impediment to productivity and innovation. Approximations indicate that overall, the digital skills gap in the UK inflicts an economic damage of over 60 billion GBP per year on the country's GDP (Department for Digital, Culture, Media and Sport, 2022). Further studies suggest that the digital and AI skills shortage in the country is underpinned with a lack of skilled workers in sustainability. Serin (2021) illuminates that the insufficiency of green skills has been identified as a challenge in meeting the targets of reducing carbon emissions by 2050. This is consistent with a report published by PWC (2022) which reveals that a large share of employees with transferable skills for the renewable energy sectors are set to retire by 2030, leaving thousands of jobs at risk of going unfilled.

Two factors that may be exacerbating the situation aforementioned are the lag in formal education and training in adapting to technological advancements (UNESCO, 2022) and the disconnect between academia and industry(Al-Sulaiti, Khalid & Dwivedi, Yogesh, 2024). Globally, education and training institutions are struggling to keep pace with AI development, resulting in curricula being misaligned from industry needs. For instance, a survey conducted by UNESCO, found that in 2022, across 190 countries only 15 were developing AI curricula in schools, and the UK was not among them. Additionally, at higher education level, where the UK has advanced AI programmes and a large percentage of graduates on STEM programmes, academics´ focus and research also differs from industry´s priorities, meaning scholar research is not always pertinent or directly applicable to the needs of the economy and the private sector (Al-Sulaiti, Khalid & Dwivedi, Yogesh, 2024).

In this context, the success of the UK in the transition towards AI and green jobs will depend on the capacity of the workforce to meet the needs of employers. Thus, understanding the skills demanded in the future of work in order to address them has become a pressing issue for policymakers, employers, and educators. Furthermore, given that this is not an isolated problem and a wide range of countries are facing similar skills mismatches, we can expect the increasing competitiveness to trigger a global competition for human capital, making the issue even more urgent.

*The shift towards skill-based hiring*

Our framework is grounded in the above mentioned theory of economics of skills, motivated by the swiftly changing labour markets due to societal and technological transitions. Firstly, we posit that human capital, specifically skills that enhance firms' profitability, correlates positively with higher income. Hence, a swift increase in the demand of skills should materialise in higher wages for jobs requiring these skills. Secondly, in situations where skills are relatively low in supply compared to high levels of demand, their market value increases and employers are compelled to adopt novel hiring practices to increase the talent pool. Applying this to the current landscape, assuming that AI and green skills contribute positively to firms' profitability and are progressively being demanded by employers while remaining relatively scarce in supply, we anticipate an increase in their market value, leading to wage premium offerings and a limited focus on candidates' qualifications for AI and green jobs as well as an adoption of skills-based hiring for jobs in these domains.

We interpret a shift to skill-based hiring for AI and green roles as a growing relevance and monetary reward for skills accompanied by a declining relevance and reward of formal education credentials. At the same time, we expect occupations with a particularly strong increase in demand for AI and green skills to show a significantly higher reward for the respective skills and a lower reward for formal degrees. Accordingly, we propose three hypotheses with regard to popularity, reward, and demand for our observation period (2018-2024) and the UK labour market:

1. *Hypothesis 1 - Popularity:*
   *The mentioning of AI and green skills increases over time (a), while the mentioning of formal education decreases for AI and green roles (b).*

2. *Hypothesis 2 - Reward:*

   *AI and green skills offer a significant monetary reward (a), while the monetary reward for formal degrees is lower for AI and green roles (b).*

3. *Hypothesis 3 - Demand:*

   *For occupations with a high demand for AI and green roles, the premium for skills is high (a), while the premium for formal education is low (b).*

**Data and methods**

To test our hypotheses, we rely on a dataset of online job vacancies, which we describe in this section. We summarise the granular-level information our data contains and its limitations. We then describe the skills-level approach used, the list of AI and green skills that served as a guideline, and the way we classified vacancies.

*Online job vacancies*

The primary data source used for this study is the online job vacancy (OJV) postings database created by Lightcast, formerly Burning Glass Technologies, and provided by the Burning Glass Institute, a labour market research non-profit. Covering the period from January 2018 to June 2024, the data encompasses a vast collection of eleven million job vacancies, with around 80 million associated skills. These listings include essential details such as job titles, job locations, employer's name and industries. Moreover, the granularity of the dataset provides information regarding the desired characteristics of candidates, including a list of skills and the educational attainment. Additionally, for the UK, the database contains data on the salaries advertised within these vacancies for about 39% of the sample. This information is not clustered across regions, industries or occupations (see Table A3 in the Appendix). AI (31%) and green (37%) postings are as likely as every other vacancy to contain wage information. Although these posted salaries may not reflect the accurate remuneration received by hired employees, they serve as an indicator of the companies' willingness to compensate for specific skill sets demanded in the job postings. A summary of the main characteristics used in this analysis can be found in Table A3 in the Appendix.

The vacancy data provides valuable and extensive insights into employer demands, but it has limitations. While we are able to analyse job vacancies, the candidate selection process is not fully transparent. Employers may consider criteria beyond those listed in postings throughout the selection process. Another important limitation is that by looking solely at vacancies posted online we may overlook other means of acquiring talent in these domains, such as external contractors or in-house skills development among existing employees. Furthermore, as highlighted by Lancaster et al. (2022), OJVs are biassed towards certain occupations, sectors and industries. Some types of vacancies are heavily represented while others are less advertised, as shown in Table A3 in the Appendix.

*A skill-based approach to identify AI and green roles*

The transition evoked by the advent of AI and a "greening" of the economy entails fundamental changes for the labour market. Yet, there is scarce systematic evidence on its impact given its complexity and the lack of a universal taxonomy and definitions. Thus, researchers have studied emerging occupations and the labour market through a wide variety of approaches. Two of the main approaches used up to date are ´bottom-up´ and ´top-down´. Top-down approaches classify jobs based on their sectors or industries (OECD, 2023). By using this approach, we would consider anyone working for a clean energy company to be working in a green job, even if their responsibilities do not involve any task related to sustainability or the environment. On the opposite side, the bottom-up approaches define jobs based on the skills or tasks they require. In this case, an employee of an AI company would only be considered to be in an AI-related job if their responsibilities involved tasks to do with AI, such as training machine learning models.

The exploration of AI implications in the labour market can be traced back to the investigations into the intersection of technology and employment. A series of studies in the 90s aimed to explain rising wage inequalities by introducing the concept of Skills Biased Technical Change (Bound and Johnson, 1992, Juhn, Murphy, and Pierce, 1993). Subsequently, researchers focused on explaining the relevance of high-skilled workers to increase productivity through the use of computers (Krueger, 1993). This trajectory of research led to new questions and academics delving into the skills that would be demanded in the workplace with the increasing use of technology and how these would affect workers with varying levels of skills (Autor, Levy, and Murnane, 2003, Manning, 2004, Autor and Dorn, 2013). However, the majority of these studies relied on a canonical model, which presented some limitations. In 2011, Autor and Acemoglu, proposed a new task-based approach that offered a better understanding of the influence of new technologies on labour and income distribution. Since then, contemporary scholars have predominantly adopted a bottom-up approach to analyse the links between AI and skills demand at a granular level (Acemoglu, & Restrepo 2019, Duch-Brown, Gomez-Herrera, Mueller-Langer and S. Tolan, 2021, Alekzeva, et al. 2021, Stephany & Teutloff, 2024 ).

The scholarly examination of green jobs and associated skills emerged later than the investigations into the implications of technology, primarily due to a historical lack of emphasis on these occupations and the complexity of identifying green jobs. Methodological constraints meant that there were no rigorous ways of identifying green jobs. Definitions were and remain varied. In 2008, the UN, for example, only defined green jobs as those which minimised waste and pollution and ILO defined green jobs as those in sectors where the carbon footprint was lower than the average. In academia, there have also been alternative ways of classifying green jobs. Echoing Sulich et al (2020) scholars were predominantly categorising green jobs through a top-bottom approach by analysing sectors like water, waste collection or renewable energies, which posed many constraints. Now, new models have emerged, many of which utilise bottom-up methodologies. In the US, O*NET codes have been used to identify green jobs (Vona et al., 2015, Dierdorff, 2011). In Korea and Germany, researchers have used data from websites that contained 'green jobs' to quantify green occupations (OECD, 2022). Saussay et al. (2022) developed a methodology to accurately isolate low-carbon activities in online job vacancy data, producing a list of 445 low-carbon skills that can be used as low-carbon job identifiers. In the UK, the Department of Business, Energy and Industrial Strategy leverages keywords and a big data approach to monitor the development of green jobs and skills (OECD, 2022). While recent bottom-up approaches still have some limitations, they serve as a good proxy to quantify jobs that require green skills and therefore, that align with human capital theory (Vona et al, 2015).

For the purpose of this study, we use a bottom-up (i.e. skill or task-based) approach to categorise AI-related and green jobs. Specifically, every vacancy that contained at least one AI skill was considered to be an AI job, and those considering at least one green skill were categorised as green, regardless of the sector. To carry out the categorisation, we used Lightcast's Open Skill Taxonomy which classifies over 30,000 unique skills. To identify AI and Green skills, we use Lightcast's up-to-date "Artificial Intelligence and Machine Learning (AI/ML)" skill subcategory which includes 157 skills and a list of green skills compiled in 2019 (Burning Glass Technologies, 2019), of which 259 skills are still present in the taxonomy. Appendix item 1 and 2 contain the lists we utilised as a guideline to identify green and AI occupations. This methodological approach enables us to not only identify the jobs that are evidently green or AI-related but also those that are less obviously so. Moreover, it also allows for comparison of skills, levels of education demanded, and salaries between occupations demanding skills in the domains of AI and green roles, and other occupations. Further, the methodology is flexible and could be easily applied across different countries and domains. An example of where information about salary, skills, or formal education is extracted from an OJV is given in Figure A5 in the Appendix.

## Results

### *Green and AI roles are on the rise*

Figure 1 summarises the main descriptive findings on the development of AI and green roles over the period from 2018 to mid-2024. In panel A1, we can see that businesses have rapidly increased the demand for employees with green and AI-related skills, as a proportion of overall job vacancies. The data shows that the number of job postings requiring at least one green skill has grown by 67% and increased by 21% AI skills between 2018 and 2023[1]. After 2023, we can see a rapid acceleration in the demand for AI skills, in particular. While green roles have constantly increased in relative demand, we see a significant drop in the demand for roles requiring AI skills in 2022, in our interpretation stemming from a period of tech layoffs in the UK during that year. To correct for changing demand due to developments in particular industries, we have normalised the demand rate by correcting for short-term occupational trends[2] (dotted line). This correction reveals a less dramatic but still significant short-term drop off in the demand for AI roles. After the 2022 layoff period, we observe a rapid rebound and steady increase in the demand for AI roles across the labour market. These trends suggest that businesses in the UK are seeking to embrace the emerging technological advances to use AI across their businesses. Further, they also suggest that employers are striving to adopt sustainable practices, or at least to meet the legal regulations regarding environmental impact.

Our findings are consistent with prior literature that shows that the impact of the green and AI roles in the labour market has become more evident in recent years, when specific events prompted and expedited the use of AI and the adoption of environmental policies. The COVID-19 pandemic of 2020, for example, increased the reliance on digital technologies (OECD, 2023) and pressured companies across all sectors to enhance their use of technology and data at a faster pace. Similarly, Russia's invasion of Ukraine in 2022 motivated many governments to boost alternative energy resources such as solar and wind energy in order to reduce gas imports from Russia (IEA, 2023).

---

[1] We define a job posting as AI-related or green if it asks for at least one of the AI and green skills provided by the taxonomy listed in our appendix.
[2] This was done by taking the average rate of AI jobs per occupation at the SOC3 level and, instead of taking the quarterly weighted average based on quarterly occupation size, weighting based on the occupations' size over the entire period, 2018Q1-2024Q2.

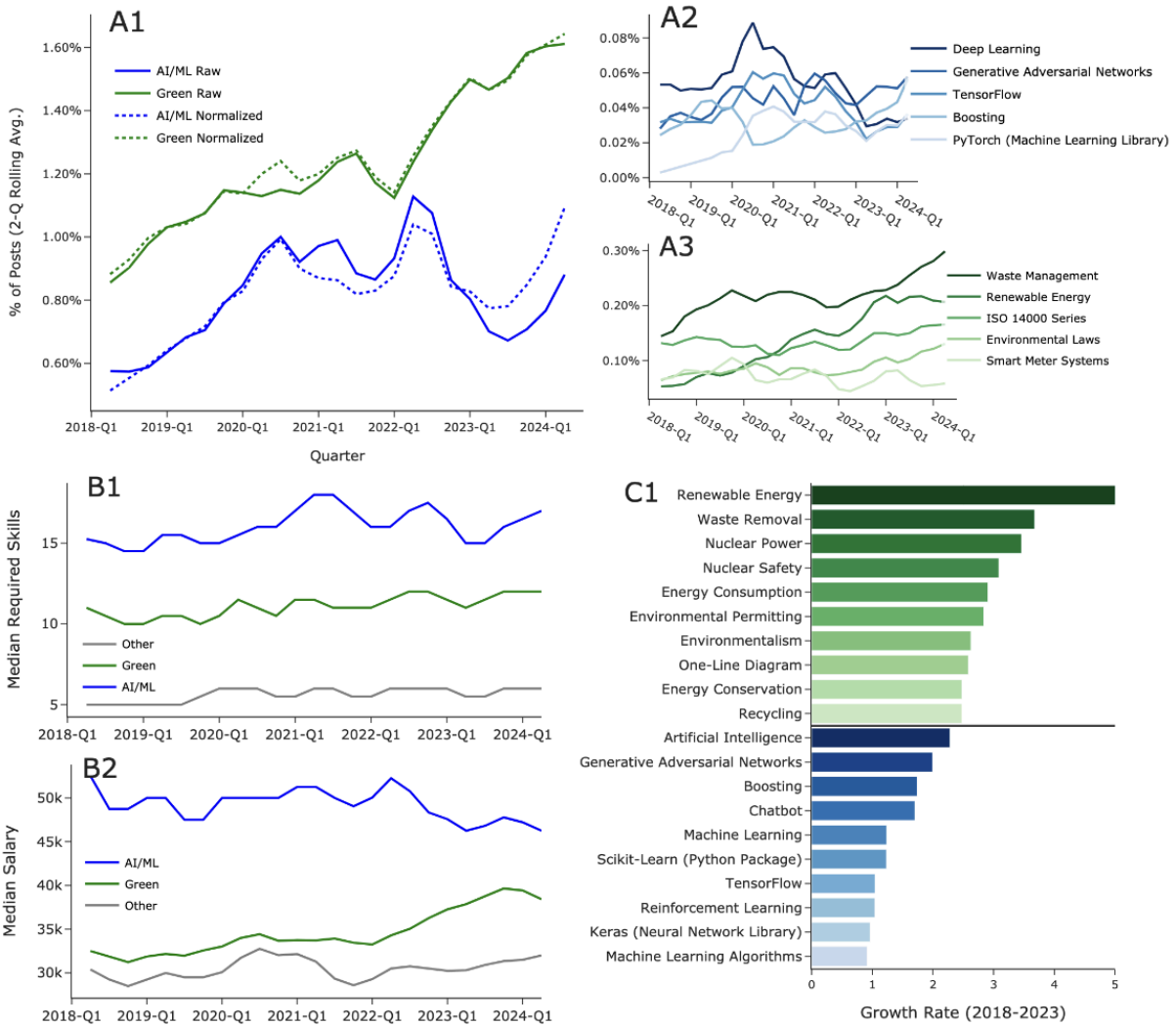

***Figure 1***. *Panel A1 shows an increasing share of job postings requiring AI and green skills. Panels A2 and A3 show the trends of the most common underlying skills which reveals that techniques like generative adversarial networks are increasingly important for AI/ML roles and skills related to waste management and renewable energies for green roles. For AI roles, the generic skills "Artificial Intelligence" and "Machine Learning" are excluded. Panel B1 highlights that AI roles are three times more skill-intensive, and green roles twice as skill-intensive, compared to the labour market average. Panel B2 illustrates that AI roles command wages around 75% higher than the average, while green roles see slightly higher wages, with growth accelerating from 2022. The strongest growth in skills is observed in renewable energy, waste removal, nuclear power, and machine learning technologies like generative adversarial networks. Panel C1 shows the fastest growing skills in each category between 2018-2023 with at least 100 related posts in 2018.*

## Which skills are needed?

The emergence of new jobs along with the disruption of existing ones will require workers to possess a different set of skills than those previously needed by employers (WEF, 2022). Our main research question revolves around the identification and quantification of the skills sought by employers for AI and green roles.

The literature suggests that potentially the main driver of change for the skills needed in the future of work is the accelerated pace at which technology is advancing. Businesses are increasingly pressured to rely on digital technologies, leading employers to look for candidates with strong digital skills. Currently, estimates show that over 80% of job vacancies in the country currently require digital skills (Department for Digital, Culture, Media and Sport, 2022). But the impact of technology does not stop there. Beyond this, machines and AI have unlocked endless possibilities to increase efficiency, reduce costs, and improve productivity (Tarafdar, 2021). Given the transformative power of AI and automation, many tasks within the labour market are increasingly exposed to forces of automation (Acemoglu, 2011). Rather than looking at the level of particular skills, this study aims to understand the *breadth* of skills demanded by firms. Recent advancements in AI have revived the idea that advancements in technology, including automation, would dramatically replace a large share of workers, putting them at high risk of unemployment. This fear is reminiscent of the current belief that the green transition will entail job displacement for workers in fossil fuel and high CO2 emission industries. However, recent research suggests that the impact of AI and the green transition on job vulnerability may not be as striking as has often been portrayed (OECD, 2023, Acemoglu, et al. 2022). Instead of jobs being completely automated, new estimates are that human–machine interaction will become ever more essential (Frey, 2019) meaning particular jobs will not disappear but instead be disrupted. Similarly, while there might be a decrease in the demand for workers in the fossil fuel extraction and processing industry, the transition towards a green economy will result in the emergence of millions of jobs (ILO, 2023). Hence, while concerns around job displacements are understandable, employment opportunities will arise, which will demand workers to fulfil non-routine tasks (World Bank, 2016).

It is evident that the skills required by employers are being reshaped. By analysing job vacancies with the lists of skills, we found that employers are looking for a wide variety of skills related to the development and implementation of AI. In Figure 1, Panel C1 highlights some of the fastest-growing skills in the domains of AI and green technologies. Skills related to generative adversarial networks (GANs), which form the backbone of advanced AI models such as ChatGPT, have shown relatively stable growth.

In the domain of green skills, Panel C1 shows the strongest growth in areas like waste management and renewable energy technologies. Panel A3 also shows there is consistent and increasing demand for Environmental Law and ISO 14000 series skills, the latter of which pertains to environmental management standards aimed at minimising negative impacts on the environment and promoting sustainability practices in industries. Panel C1 further emphasises that the growth in individual green skills generally outpaces that of AI skills over this time period. Renewable energy skills and waste removal expertise have seen particularly sharp increases, with renewable energy-related skills growing fivefold from 2018 to 2023, and nuclear power and safety skills also tripling in the same period. In the AI field, there has been a strong rise in general AI skills, which more than doubled over the observation period, along with notable growth in skills centred around generative adversarial networks and chatbot development.

Panel B1 provides insights into the skill intensity of AI and green roles, highlighting the number of skills typically required in these positions. While the average job posting in the broader labour market demands around five skills, this number doubles for green roles, where approximately 10 skills are required. For AI roles, the demand for skills is even higher, with an average of 15 to 16 skills per position. This suggests that both green and AI roles are more skill-intensive than average, underscoring the importance of a diverse skill set in these rapidly growing fields.

Panel B2 shows a clear difference in advertised wages between AI, Green, and other postings. While the average salary for job postings remains slightly below £30k, green roles tend to offer higher salaries, with an upward trend reaching towards £40k by the end of the time series. AI roles, on the other hand, command significantly higher salaries, averaging around £50k. However, there is a slight decline in AI salaries during the period from early 2022 to mid-2023, which corresponds with a reduction in demand due to tech sector layoffs. Despite this dip, AI roles remain among the highest-paying positions.

*Does education still matter for green and AI roles?*

Traditionally, employers have relied on formal education qualifications as a way of identifying potential employees. Thus, higher educational credentials have tended to be positively associated with employment rates and earning advantages (Mincer, 1974, Carbonaro, 2007; OECD, 2022). However, in recent years, this approach may have become less effective, as a large number of adults are acquiring their skills and experience through non-formal or informal education. Non-formal education refers to structured learning that takes place outside the traditional classroom environment, such as on-the-job training or apprenticeships, and informal education consists of unstructured learning, which takes place in everyday experiences, such as learning a new skill through a hobby or interest.

The dominance of non-formal and informal education among the working age population means that many skills are not necessarily reflected in the workforce´s formal qualifications (OECD, 2021). As a result, employers who focus exclusively on formal qualifications may be overlooking capable candidates who have gained human capital through non-traditional routes. To address this issue, many employers have reported that they are increasingly adopting a skills-based approach to recruitment, focusing on the specific skills of candidates, rather than relying solely on higher education degrees. For example, formal education is not mentioned in a posting or not requested as compulsory (Burning Glass Institute, 2022). Instead a precise set of skills is outlined and how these competencies contribute to success in the advertised position.

The correlation between education and employment rates has remained steady over recent decades. However, research shows that the significance of education credentials in the eyes of employers is influenced by external factors such as social shifts and economic disturbances (Burning Glass, 2022, Fuller & Raman, 2017). Some authors suggest that there has been a growing vertical demand for higher education credentials, as a result of the accelerated growth of educational opportunities and the proliferation of undergraduate degrees (Collins, 2019). Similarly, it has been found that in times where the number of vacancies is scarce compared to the pool of qualified candidates, employers are likely to add qualification requirements as the supply of educated candidates exceeds demand (Brown and Souto-Otero, 2020, Golding & Katz, 2008). In labour economics and sociology, this trend of adding a minimum level of education to jobs that would previously not require them, is referred to as ´degree inflation´ (Brown and Souto-Otero 2020). This trend was particularly visible during the economic disturbance between 2007 and 2010. Throughout the 2008 financial crisis, job postings demanding a bachelor's degree as a minimum level of education increased by more than 10% (Modestino et al., 2015, Burning Glass Institute, 2022).

On the opposite side of degree inflation, research suggests that skills shortages tend to encourage employers to seek an alternative to recruiting workers, namely skills-based hiring. For example, during the COVID-19 pandemic, where there was a significant lack of skilled medical workers, the number of job postings for medical staff requiring minimum education levels decreased (Fuller et al., 2022, Burning Glass Institute, 2022). By eliminating the minimum level of education from job postings, employers increase the talent pool and may be able to find more suitable candidates. The main focus of this article is to understand whether a similar phenomenon is now occurring with jobs demanding AI or green skills, for which employers are struggling to find the right candidates and which don´t yet have clear career paths.

Focusing on formal education as a currency for employability opportunities seems to be insufficient. Several formal education degrees related to the domain of AI and green roles have been established only very recently in leading higher education institutions around the world (Chen et al., 2020; Muñoz-Rodríguez et al., 2020). Beyond education credentials, green and AI roles also require a broad set of skills. Whilst education qualifications signal that individuals have become highly educated in certain domains (Becker, 1975), it may be the possession of specific skills and experience that will truly distinguish candidates. By enlisting skills in their job postings, employers can effectively identify candidates with a specific combination of technical and soft skills, ensuring a higher likelihood of successful job placements (Stephany & Teutloff, 2024).

Including skills on job postings allows for consideration of skills developed outside formal education (Fuller, 2022) which currently represents the predominant way of learning among workers through non-formal and informal education. Data indicates that over 70% of workers learn through their job and interactions with others, compared to only 8% who are enrolled in formal education programmes (OECD, 2020). This is particularly relevant for technological skills, as they tend to be acquired on the job or through informal networks (Collins, 2019).

The skills demanded by firms have been the subject of extensive research. Numerous studies have analysed the demand for skills, the wage premium related to specific competencies (Stephany & Teutloff, 2024), and the correlation between skills and employment opportunities (see Acemoglu, 2011; Gofman, 2020; Deming, 2018). Many of these studies have found that as technology becomes pervasive across industries, developed economies are increasingly requesting more highly skilled workers, while reducing the number of low skilled and routine jobs (Autor, 2014, Goos et al., 2014).

While supply and demand may influence the minimum levels of education demanded by employers, the nature of occupations can also lead employers to demand diverse education credentials. More specialised and complex occupations tend to require higher education degrees. Given the specialised nature of green and AI-related occupations, jobs in these fields are likely to require highly educated workers. In our data we find that the pertinence of formal education for green and AI jobs has changed significantly in the last six years, as summarised in Figure 2.

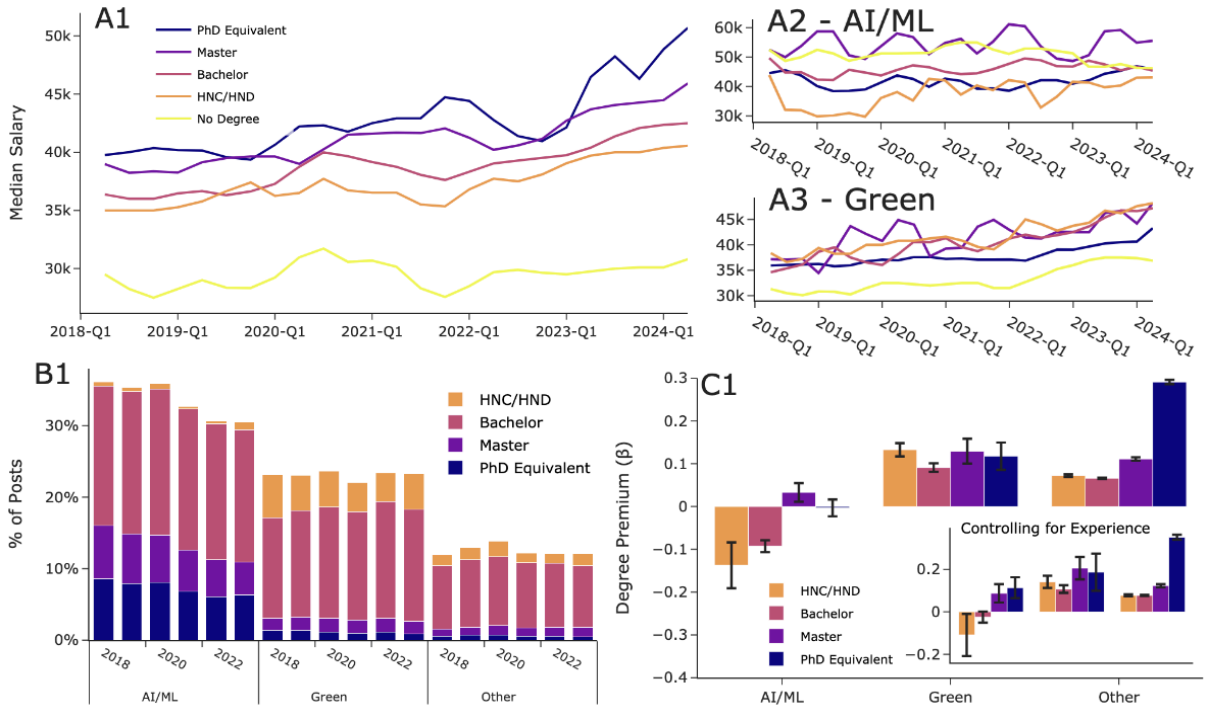

***Figure 2***: *Panel A1 illustrates the wage premium associated with higher education across all roles, with PhD holders earning the highest salaries, reaching around £50k by the end of the time series, while job postings without degree requirements offer significantly lower wages around £25k. Panel A2 shows a different trend for AI roles, where the highest-paying jobs are associated with master's degrees, but wages for roles with no formal education requirements are comparable, and there is no significant wage premium for PhDs. Panel A3 reflects a more consistent educational gradient for green roles, where higher education correlates with higher wages, although there is also a notable increase in wages for roles without formal degree requirements. Panel B1 highlights the higher educational requirements for AI and green roles, with 23% of green roles and 33% of AI roles demanding formal education, compared to just 12% of all other roles. Finally, Panel C1 shows a clear educational wage gradient in the broader labour market, with technical, bachelor's, master's, and PhD degrees offering significant premiums. However, this gradient is flattened for green roles and nearly absent for AI roles, indicating a shift in reward structures in these sectors. Full regression results for C1 can be seen in Table 1 of the Appendix*

Figure 2 provides key insights into the relationship between education and salaries, particularly in green and AI roles. Education is often seen as a driver of higher salaries, and Panel A1 supports this, showing significantly higher wages for roles requiring formal education, especially PhDs, which includes professional doctorates. Job postings that do not require a degree hover around £25k, whereas positions requiring technical qualifications (HNC, HND), bachelor's, master's, or PhD degrees all offer salaries above £35k, with PhD roles nearing £50k towards the end of the time series.

Interestingly, this trend remains largely consistent for green roles in Panel A3. Higher education correlates with higher wages, but there is also a noticeable increase in salaries for roles with no degree requirement, suggesting that growing demand in the green sector has pushed up wages across the board, even for less qualified roles. Notably, PhD roles in the green sector do not significantly outperform bachelor's, master's, or technical degrees, indicating less of a premium on advanced degrees in this domain. This could be, in part, due to many PhDs in this category remaining in academic positions which typically pay less than industry.

In contrast, Panel A2 paints a different picture for AI roles. Here, the salary structure appears less straightforward. The highest-paying positions tend to require a master's degree, but roles with no formal degree requirement also offer relatively high wages, comparable to those with technical, bachelor's, or PhD qualifications. This suggests that the reward structures in AI have shifted, with education playing a less decisive role in determining salary levels compared to other factors. This aligns with contemporary discussions on the changing nature of AI jobs. For example, Autor (2024) suggests that AI brings tasks previously performed by high-skilled experts within reach of middle-skilled individuals, potentially reducing the premium for higher education. Our findings align with this hypothesis, as we observe that in AI roles, the premium for master's and Ph.D. degrees has diminished. This indicates that AI may be enabling individuals without advanced degrees to perform complex tasks, thereby reducing the wage premium traditionally associated with higher education.

Panel B1 highlights the education intensity of green and AI roles compared to the broader labour market. While only around 12% of general job postings require formal education, 23% of green roles and nearly 33% of AI roles specify educational requirements. Additionally, green roles show a higher demand for technical qualifications (HNC, HND), reflecting the sector's technical nature, while AI roles demand a higher share of master's and PhD degrees. However, there's a notable decline in formal education requirements for AI roles over time: in 2018, 36% of AI roles required formal education, but this dropped to around 31% by 2023. This suggests a shift towards valuing skills and experience over formal education in AI roles.

Reflecting on the findings summarised in Figure 1 and 2, we find evidence in support of our first hypothesis. For one, we clearly observe that both AI and green skills have become more demanded in our observation period. At the same time, at least for AI roles, we observe that the mentioning of formal degree requirements has decreased visibly.

*What is the premium of AI and green skills?*

We aim to further understand and compare the relationship between the minimum level of education and the skills demanded with the remunerative outcomes. Generally, jobs that demand higher educational attainment tend to have sizable wage premiums. However, earning inequalities are not solely based on educational attainment; field of study and the type of occupation are also relevant factors that influence economic rewards (Carbonaro, 2007).

Previous literature has posited a strong correlation between economic compensation, and the dynamics of skills availability and labour supply and demand (Autor, 2014). Consequently, wage premiums have been recognised as valuable indicators of potential skill shortages, recruitment challenges, and the value of specific skills (Stephany & Teutloff, 2024; Saussay et al., 2022). Prior studies on green and AI occupations have shown that these jobs offer higher economic rewards (Vona et al., 2019; Alekseva et al., 2021). This tendency in wage premia is consistent with the skills shortages reported by employers across these domains.

The novelty of this study is to delve into the variations in wages among different job vacancies and estimate the wage premium associated with AI and green skills. If jobs requiring AI skills are expected to yield higher productivity due to the application of these skills, it follows that such vacancies will offer a significant wage premium. Furthermore, we anticipate that the AI and green skills premium may differ across skill types. For instance, skills such as Generative Adversarial Networks or Waste Management are in high demand and should therefore yield a higher economic return.

On average, we see that wages in AI and green OJV postings are higher than in other job advertisements (see Figure 1-B2). In contrast to the premium of higher formal education across all postings (Figure 2-A1), we see that AI roles offer wages on a stable level and significantly higher than any role mentioning formal education can offer on average. However, it is worth noticing, as a limitation of this work, that we only observe offered salaries by the nature of the data. This means that the later negotiated salaries for the position advertised might differ from the wage levels considered in our analysis. While the salaries listed may not accurately reflect the actual compensation received by employed individuals, they serve as a reliable indicator of the company's willingness to pay for the specific skill set required for the vacant position.

To further investigate the changing relevance of formal education for AI and green roles, we conducted a regression analysis explaining differences in offered wages. To control for potential confounders, such as industry type, occupational sector, or regional disparities, we run a regression model to explain wage premia for formal education across the AI and green roles:

$$log(wage_i) \ = \ \beta_0 \ + \ \beta_1 \ * \ skill_i \ + \ \beta_2 \ * \ edu_i + \ \beta_3 \ * \ exp_i + \eta_i \ + \ \epsilon_i \qquad (1)$$

The regression model explains the logarithmic asking wage of each posting (i) with a dummy for whether the post mentions at least one AI or green skill ($skill_i$), a set of dummies for the level of education ($edu_i$), the required years of experience ($exp_i$), and fixed effects ($\eta$) for four years, twelve regions (NUTS 1), 16 industries (SIC 1), and 9 occupational groups (SOC 1).

Table 1 summarises our regression model explaining offering wages. Required experience and the fixed effects are included in our model at all times. In addition, we variably include educational controls and dummies for whether a posting has any AI or green skill. In our model, an educational gradient is visible where higher educational attainment is associated with higher offering wages, as shown in models two, four and six. Similarly, we observe a premium for AI skills in models three and four. However, green skills, which are in less demand than AI skills, do not show a large wage premium, as shown in models five and six.

**Table 1**. *An educational gradient is evident (Models 2, 4, 6), with higher education linked to higher wages. AI skills show a significant wage premium of 23%, second only to holding a PhD (33%) and higher than a master's degree (13%) (Models 3 and 4). Green skills do not exhibit a significant wage premium (Models 4 and 5).*

| | (1) | (2) | (3) | (4) | (5) | (6) |
|---|---|---|---|---|---|---|
| Dep. Var. | | | | *log(salary)* | | |
| **Variables** | | | | | | |
| *ai* | | | 0.26*** | 0.23*** | | |
| | | | (0.00) | (0.00) | | |
| *green* | | | | | 0.04*** | 0.03*** |
| | | | | | (0.00) | (0.00) |
| *hnc/hnd* | | 0.08*** | | 0.08*** | | 0.08*** |
| | | (0.00) | | (0.00) | | (0.00) |
| *bachelor* | | 0.08*** | | 0.08*** | | 0.08*** |
| | | (0.00) | | (0.00) | | (0.00) |
| *master* | | 0.13*** | | 0.13*** | | 0.13*** |
| | | (0.00) | | (0.00) | | (0.00) |
| *phd equivalent* | | 0.35*** | | 0.33*** | | 0.35*** |
| | | (0.01) | | (0.01) | | (0.01) |
| *experience* | 0.01*** | 0.01*** | 0.01*** | 0.01*** | 0.01*** | 0.01*** |
| | (0.00) | (0.00) | (0.00) | (0.00) | (0.00) | (0.00) |
| *intercept* | -50.89*** | -51.05*** | -50.80*** | -50.97*** | -50.78*** | -50.96*** |
| | (0.0) | (0.0) | (0.0) | (0.0) | (0.0) | (0.0) |
| **Fixed Effects** | | | | | | |
| *year* | yes | yes | yes | yes | yes | yes |
| *sic 1* | yes | yes | yes | yes | yes | yes |
| *soc 1* | yes | yes | yes | yes | yes | yes |
| *nuts 1* | yes | yes | yes | yes | yes | yes |
| R-squared | 0.33 | 0.33 | 0.33 | 0.33 | 0.32 | 0.33 |
| R-squared Adj. | 0.33 | 0.33 | 0.33 | 0.33 | 0.32 | 0.33 |
| N | 899160 | 899160 | 899160 | 899160 | 899160 | 899160 |

Standard errors in parentheses, * $p < .1$, ** $p < .05$, *** $p < .01$

Overall, our findings indicate a shift from the usual focus on educational attainment as a means to a wage premium. For example, even when controlling for geographical, sectorial, occupational, and educational variance, AI skills exhibit a wage premium on offering wages of 23% - second only to having a PhD (33%) - and significantly higher than for a master's degree (13%). The study shows that the minimum level of education required within AI jobs wields diminished influence over the wage premium, which supports the idea that firms are applying skills-based hiring practices within their recruitment processes for AI jobs. Accordingly, this last part of our analysis delivers strong evidence in favour of the first part of our second hypothesis—that there is a premium in offering wages for skills in AI and green domains. For the case of AI roles, the skill premium is comparable to formal degrees.

*Is the degree premium lower for AI and green roles?*

We find clear evidence that the skill premium is comparable to the economic reward of formal degrees for AI and green roles. But how about the premium of education within these roles? Did the premium on formal degrees diminish for AI and green roles, as postulated in our hypothesis 2b? Here, we run the above-mentioned regression model (1) for the subsample of AI and green roles respectively. The findings on the premium of formal degrees are summarised in Figure 2C1 and the full results can be found in Table A1 in the Appendix. For job offers outside of AI and green roles, Panel C1 illustrates a clear so-called educational gradient (rising premium the higher the degree). The regression results show a significant wage premium for those with technical or bachelor's degrees, an even higher premium for master's degrees, and a substantially greater premium for PhD holders—three times higher than for those with master's degrees. These coefficients are derived from a model where the reference group is job postings with no education requirement.

However, for green roles, the educational gradient appears to have flattened. While roles requiring formal degrees still offer higher wages, the difference between educational levels—such as technical, bachelor's, and master's degrees—is not as pronounced. Interestingly, technical qualifications like HNC and HND are on par with traditional formal education in terms of wage returns. The most striking results come from the AI roles. The regression shows that the wage premium for formal education has nearly disappeared. There is only a slight premium for master's degrees compared to no education, and no significant premium at all for PhDs. In fact, for AI roles, the beta coefficient is negative for both technical qualifications and bachelor's degrees, indicating that AI roles that don't require a degree are typically seeking workers with more skills than these degrees offer. Therefore these degrees may not lead to higher wages in this field. This suggests that the demand for skills in AI has shifted away from formal educational qualifications, with a greater emphasis on practical skills and experience.

When controlling for experience, as seen in the subplot of the figure, the educational gradient reappears slightly but remains much weaker for AI and green roles compared to other sectors. This confirms our hypothesis that in AI and green sectors, the monetary rewards associated with specific skills and experience are similar to those typically attributed to formal education. In contrast, the wage premium for formal degrees, particularly for higher-level degrees like master's and PhDs, is significantly lower for AI and green roles compared to the rest of the sample.

The findings of this analysis support our hypothesis 2b in part. Clearly, for AI and green roles, higher level degrees, in particular PhDs, have lower premia than in other job postings. In the case of green roles, formal education is still associated with higher wages overall, but the educational gradient has flattened. For AI roles, only when controlling for experience do we observe a slightly positive premium for master and doctoral degrees.

So far, we have defined green and AI roles as postings that require at least one of the respective skills. But how does the phenomenon develop once we increase this requirement to more than one skill, that is examining the compounding effects of skills? As Figure A1 in the Appendix illustrates, increasing the threshold from one to two AI/green skills reduces the sample size significantly. Nonetheless, we examine our two main metrics – skills' and educational premium – while increasing the threshold of how we define AI and green roles respectively, as shown in Figure A2. Here, we see that for AI roles, the AI premium clearly increases the more skill intensive a role becomes, while the return on formal degrees diminishes. For green roles, the return on skills is relatively stable and the economic reward offered for formal degrees decreases slightly. Both findings indicate that the phenomenon of skill-based hiring is even more pronounced for roles that require multiple skills.

*Does higher demand lead to skill-based hiring?*

Finally, we examine to what extent the demand for AI and green skills is actually related to a wage premium for these skills, in addition to education. For this analysis we iteratively run alternations of regression model 1 (after removing the occupational fixed effect), presented in the previous section, for subsamples of 26 occupations (SOC 2) and six years, from 2018 to 2023. Accordingly, we interpret the models' beta coefficients for the AI and green skills dummy as a wage premium. Additionally, over the same set of samples, we run alternations of the model, replacing the skill dummy with a constructed degree dummy which indicates whether a post requires a bachelor's degree or higher (after removing the education control). Our third hypothesis assumes that the wage premium for skills should increase with demand while the degree premium should decrease. The relationship between these characteristics of interest is shown in Figure 3C1-C4.

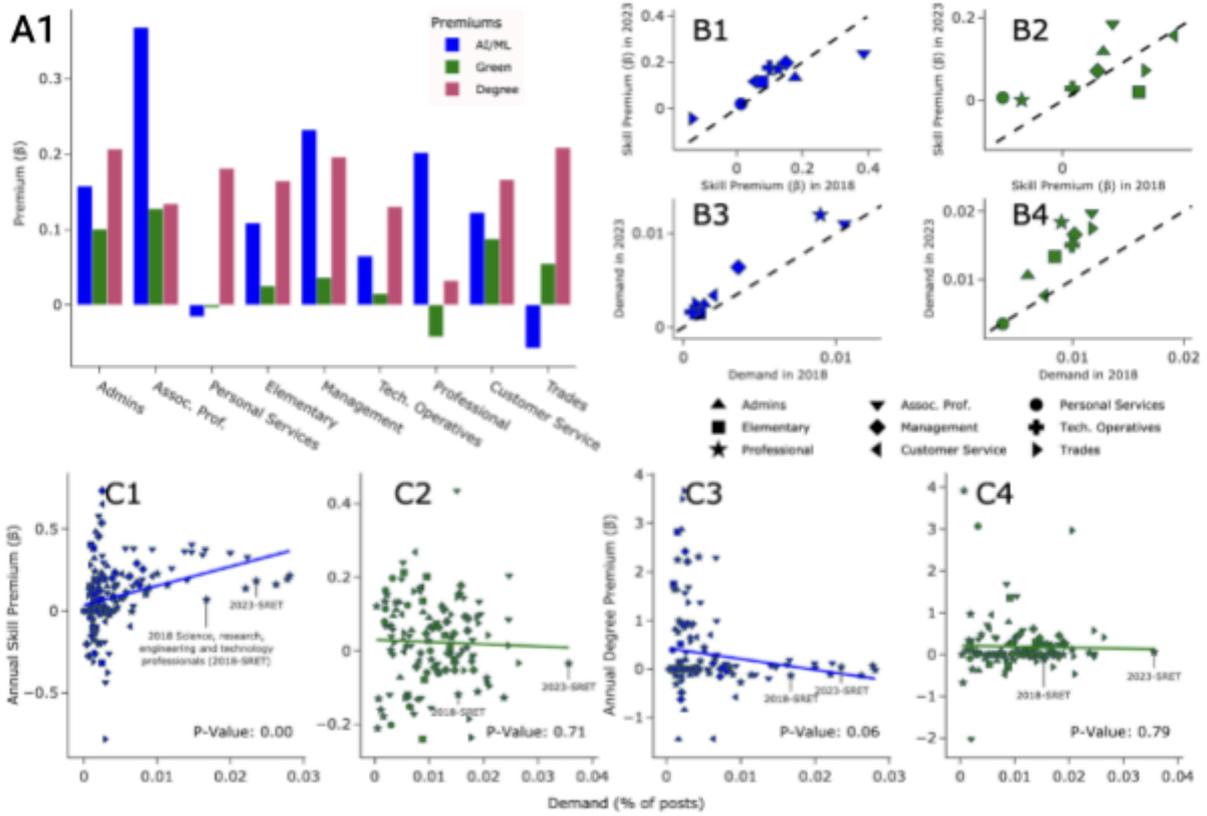

***Figure 3***. *(A1) Shows the AI, Green, and Degree premiums at the SOC1 level controlling for experience, year, SIC1, and NUTS1. For Professionals and Management jobs, AI skills have a significantly higher premium than degrees. More granular results at the SOC2 level can be found in Table A4 in the appendix (B1-2) Shows the change in skill premiums for AI and green skills, respectively, for each of the SOC1 occupations between 2018-2023. (B3-B4) Shows the change in demand as a percent of total postings within each of these occupational groups. (C1-C2) Show the relationship between the skill premium and demand across years and SOC-2 occupations for AI and Green skills. (C3-C4) Show the relationship between the degree dummy premium. For the case of AI skills, we observe a clear positive relationship between skill demand and skill premium and a negative correlation between demand and degree premium.*

Figure 3A1 shows that skill and degree premiums are not equally distributed across occupations. For example, the AI skill premium is significantly higher than the offered reward for degrees in occupations related to Associate Professionals and management jobs, while in Trades and Professional Services degrees still offer a higher reward. In Figure 3B3, we can see that across SOC1 occupational groups there isn't a significant increase in demand for AI skills over the timespan of our sample, like there is for green skills. Nevertheless, C1 and C3 in Figure 3, support our hypothesis that skill premiums increase with, and degree premiums decrease with, demand for AI skills. This effect is primarily driven by variation across occupations. There are a handful of Professional and Associate Professional occupations that have unusually high demand for AI skills, high skill premiums, and low degree premiums. This indicates that employers looking for AI skills are already applying skill-based hiring while this doesn't seem to be the case for Green skills. In a second step we explain these year (i) and occupation (j) specific premiums with a set of characteristics, namely, year and occupations fixed effects, occupation size (count), skill demand, and change in skill demand over the previous two-years[3], that is, the change in the share of postings requesting AI or green skills in the respective group.

$$premium_{i,j} = \beta_0 + \beta_1 * year_i + \beta_2 * occ_j + \beta_3 * count_j + \beta_4 * (\Delta)\, demand_j + \epsilon_i \quad (2)$$

Our regression model confirms the findings from Figures 2C1-4. We see that the share of postings requesting an AI skill is positively related with the wage premium in the occupation. In other words, the more AI skills are demanded within an occupational field, the higher the associated wage premium, as summarised in Table 2.

**Table 2**. When explaining the occupational wage premium, we consider variation across time (four years), type of skill (AI or green), size of the occupation (number of OJVs, log), and skill demand (share of OJVs requesting AI or green skills). For both AI and green skills, our model shows that wage premium for a respective skill increases with the demand for it.

| | AI/ML | | | | Green | | | |
| | (1) | (2) | (3) | (4) | (5) | (6) | (7) | (8) |
|---|---|---|---|---|---|---|---|---|
| Dep. Var. | Skill $\beta$ | Skill $\beta$ | Degree $\beta$ | Degree $\beta$ | Skill $\beta$ | Skill $\beta$ | Degree $\beta$ | Degree $\beta$ |
| **Variables** | | | | | | | | |
| *demand* | 10.08*** | | -30.85* | | 1.21 | | 2.65 | 2.65 |
| | (3.45) | | (18.53) | | (1.07) | | (8.35) | (8.35) |
| $\Delta$ *demand* | | -0.01 | | -0.10 | | 0.00 | | |
| | | (0.02) | | (0.10) | | (0.00) | | |
| *log(count)* | 0.00 | 0.02 | -0.04 | -0.15 | 0.03*** | 0.03*** | -0.14** | -0.14** |
| | (0.02) | (0.02) | (0.12) | (0.11) | (0.01) | (0.01) | (0.06) | (0.06) |
| *intercept* | 0.08 | -0.08 | 1.00 | 2.21* | -0.29*** | -0.29*** | 1.47** | 1.47** |
| | (0.19) | (0.19) | (1.29) | (1.21) | (0.11) | (0.11) | (0.60) | (0.60) |
| **Fixed Effects** | | | | | | | | |
| *year* | yes | yes | yes | yes | yes | yes | yes | yes |
| *soc1* | yes | yes | yes | yes | yes | yes | yes | yes |
| $r^2$ | 0.30 | 0.24 | 0.13 | 0.11 | 0.20 | 0.19 | 0.18 | 0.18 |
| $r^2$ Adj. | 0.20 | 0.13 | -0.02 | -0.04 | 0.08 | 0.08 | 0.05 | 0.05 |
| N | 103 | 103 | 93 | 93 | 104 | 104 | 100 | 100 |

Standard errors in parentheses, * $p < .1$, ** $p < .05$, ***$p < .01$

---

[3] We used a two-year lag variable as employers will not immediately respond to a tight labour market by increasing offered wages

*Summary: Skill-based hiring for AI roles, mixed results for green jobs*

In summary, the findings indicate that our three hypotheses were, to certain extent, substantiated. Firstly, regarding a change in relevance, our quantitative analysis shows an increase in the overall mention of AI and green skills. At the same time, at least for AI roles, the emphasis on formal education has decreased. Over the past four years, job postings requiring at least one AI skill have increased by nearly 21% and increased by 67% for green skills. In addition, for AI roles, the mentioning of formal higher education has decreased by 15% from 2018 to 2023.

Secondly, regarding monetary rewards, our analysis revealed that OJVs asking for a formal degree offer wages higher than postings without higher education requirements - 8% higher for a bachelors, 13% for a masters and 35% for a PhD degree. This educational gradient of reward is much flatter for green OJVs. However, for AI postings, the higher education premium has disappeared, as postings in the field of AI offer comparable wages, whether they ask for a university degree or not, showing that candidates´ AI skills and experience are as – or even more – valuable for firms than university degrees. In addition, our regression analysis shows that AI skills offer a monetary reward on par with formal education requirements, as they exhibit a wage premium of 23% which sits between a masters degree and PhD.

Lastly, we are able to relate the demand for skills with their wage premium, in addition to education. Here, again, AI skills and green skills have different properties. For AI skills, we observe that wage premiums increase with demand, as hypothesised. This implies that for occupations where AI skills are in particularly high demand, skills-based hiring practices have been adopted. For green skills, however, the relationship between demand and premium is flat.

## Conclusion and Policy Recommendations

Our work sheds light on the increasing demand for AI and green skills, and also on the specific requirements imposed by employers and the wage differences and premia associated with these skills. The findings show important trends and disparities between AI-related and green jobs, providing valuable insights for policymakers, educators, employers and job seekers.

Our analysis of eleven million OJVs reveals a rising demand for AI and green skills in the UK, reflecting the continuous efforts of firms to adopt AI – and the AI-human complementarity– and meet environmental sustainability targets. Our analysis further suggests that employers are seeking AI and green skills differently, reducing degree requirements for AI roles while increasing wages for green roles. . Moreover, AI-related occupations typically require higher levels of education and a larger number of skills, reflecting the complex nature of these roles. The initially high rate of degree requirements and need for highly skilled workers indicate that employers looking to fill AI roles have a lot to gain by introducing skills-based hiring practices. We found evidence for these practices in occupations with high demand for AI skills which offer a significantly augmented remuneration for highly-skilled workers regardless of their educational background. Job postings demanding green skills were found to possess more similar education requirements compared with the average of other roles. This may reflect the greater variation in green roles which can span dissimilar occupations such as environmental engineers and climate policy analysts. While they still demand more skills and education than average, the requirements are less stringent compared to AI-related jobs. In line with that, while green vacancies still offer competitive remunerations, they have a less striking wage premium compared with other job postings. This may imply that up-to-date firms have started adopting a skills-based hiring approach for AI jobs, however, this shift in hiring practices has not yet happened, in general, for jobs requiring green skills.

Addressing our research question, we find indications that UK firms have indeed switched to a practice of skill-based hiring for occupations that demand AI skills. We see that the demand for AI and green positions has increased in the six years of our observation period. We show that individual skills, rather than formal education requirements, have become increasingly important features of job advertisements for AI roles. In addition, our analysis shows that individual skills, in addition to educational attainment, help to explain the significant salary premium offered by AI roles. However, the increasing gradient for educational premium - that is, offering higher wages with higher university degrees - has disappeared for AI roles. Rather skills, like generative adversarial networks or building chatbots with systems like ChatGPT, explain higher wage levels across industries and occupations.

Based on the findings of our analysis and previous policies and practices, several recommendations can be made to address these novel skills demands. The labour market is dynamic and will continue to change. To better understand and address the changes, some organisations and policymakers, such as CEDEFOP, have started developing quantitative projections of expected trends in employment (CEDEFOP, 2022), which are complementary to national estimates. Forecasting skills demands plays a crucial role in fostering a proactive approach to monitoring and identifying new job types and the skills these will require. While efforts have started being made in this direction, policymakers, academia, and employers should increasingly work together to recognise emerging occupations and anticipate employers' needs. This would allow relevant stakeholders to make better informed decisions and adapt their upskilling and reskilling strategies accordingly. An effective way of analysing the labour market is by examining online job vacancy data. As demonstrated in this paper, OJV data contains granular-level characteristics of job postings, which can provide valuable insights for key stakeholders. Governments can leverage OJV to align their educational, employment, economic, and migration policies to the needs of the local labour market. Similarly, by identifying skills demands, firms can support their employees by providing opportunities that equip them with the skills needed to remain competitive.

As mentioned above, the literature suggests that the demand for green and AI skills is far exceeding supply, making it hard for employers to find candidates with the skills needed to navigate societal and technological change. Therefore, it has become a pressing issue for the public and private sector to future-proof the workforce's skills. In order to ensure the workforce acquires the skills needed by the economy, it is essential to align and adapt education and training systems to emerging occupations. Given the weight placed by employers on university degrees, a good avenue for doing so might be through higher education. However, relying solely on higher education is likely to be insufficient as it disregards a significant portion of the working-age population and presents many challenges, such as overlooking skills acquired outside formal education, limiting the access to employability opportunities, and exacerbating socio-economic disparities. To ensure the adequacy of human capital to the current green and AI transitions, employers and governments should mainstream alternative skills development initiatives. There is a wide variety of avenues that could be implemented to develop skills outside of higher education, including apprenticeships, on-the-job training, MOOCs, vocational education and training, and bootcamps.

Some countries and employers have started leveraging these alternatives to boost the development of digital and green skills. The Netherlands, for example, currently offers workers over 45 with guidance on their job, as well as grants for training to SMEs in the ICT sectors, digitalisation or the green transition. Similarly, the Swedish government is currently providing older adults with digital skills training through the Digidel network and centres, which offer courses on digital communication and digital security (OECD, 2023). In parallel, many employers across the US and the UK, including Verizon and Citi, are providing their employees with the opportunity to develop data and digital skills through apprenticeships (Chopra-McGowan 2021), and many others are offering in-house training on how to leverage novel technologies.

Lastly, employers could benefit from resetting their hiring and talent management approaches. Up-to-date stakeholders from the private and public sectors have started this transformation, leading to the increasing prominence of skills-based recruitment. For example, the State of Maryland has removed the four-year college degree requirement for many state jobs, in order to promote inclusivity for individuals who have acquired skills through alternative paths. In the private sector, IBM is leading the way in modernising hiring approaches by making over 50% of their job openings in the US skills-based, rather than contingent on a traditional four-year degree, thus fostering a more open and inclusive job market (IBM, 2022). Given that education and industry will keep evolving at uneven paces, employers should progressively focus on candidates' skills rather than on their formal qualifications.

By requesting a specific level of education, employers may limit the number of candidates and overlook individuals who have developed skills through alternative paths such as on-the-job learning, online courses and social interactions. A skills-based hiring approach can increase the number of potential candidates, the variety of workers' social backgrounds and add diverse insights to the workforce. Additionally, in emerging fields such as AI and the green domains, where employers are struggling to find the right talent, attracting and recruiting candidates based on skills rather than formal education degrees may contribute to increasing the size of the talent pool and potentially, tackling skills shortages.

# Appendix

## Appendix 1 - List of skills used to identify AI skills

*Artificial Intelligence and Machine Learning (AI/ML) Subcategory*

'Artificial Intelligence', 'Reinforcement Learning', 'Voice User Interface', 'Machine Learning Model Training', 'Deep Learning Methods', 'Computational Intelligence', 'Dialog Systems', 'Transformer (Machine Learning Model)', 'Intelligent Systems', 'Scikit-Learn (Python Package)', 'Knowledge-Based Configuration', 'AIOps (Artificial Intelligence For IT Operations)', 'Language Model', 'AdaBoost (Adaptive Boosting)', 'OpenAI Gym', 'Dlib (C++ Library)', 'Google Cloud ML Engine', 'PyTorch Lightning', 'mlpack (C++ Library)', 'Generative Adversarial Networks', 'Recommender Systems', 'MLOps (Machine Learning Operations)', 'Knowledge Engineering', 'OpenCV', 'Theano (Software)', 'Open Neural Network Exchange (ONNX)', 'Intelligent Control', 'Text-To-Speech', 'Attention Mechanisms', 'Game Ai', 'H2O.ai', 'AI Copywriting', 'Adversarial Machine Learning', 'OpenVINO', 'Pydata', 'Seq2Seq', 'Google Bard', 'IPSoft Amelia', 'Apache SINGA', 'Caffe (Framework)', 'Chatbot', 'Apache Mahout', 'Dask (Software)', 'Keras (Neural Network Library)', 'Autoencoders', 'Long Short-Term Memory (LSTM)', 'Azure Cognitive Services', 'AI/ML Inference', 'Applications Of Artificial Intelligence', 'Cognitive Computing', 'Amazon Alexa', 'Watson Studio', 'Explainable AI (XAI)', 'Programmatic Media Buying', 'Generative Artificial Intelligence', 'Cognitive Robotics', 'Bot Framework', 'Kubeflow', 'Fast.ai', 'AWS Certified Machine Learning Specialty', 'Semi-Supervised Learning', 'PaddlePaddle', 'Meta Learning', 'Google AutoML', 'Feature Learning', 'Weka', 'Transfer Learning', 'Swarm Intelligence', 'Watson Conversation', 'Prompt Engineering', 'Stable Diffusion', 'TensorFlow', 'Boosting', 'AWS SageMaker', 'Machine Learning Methods', 'Feature Extraction', 'Caffe2', 'Feature Selection', 'Training Datasets', 'Artificial Intelligence Markup Language (AIML)', 'Intelligent Agent', 'Gesture Recognition', 'PyTorch (Machine Learning Library)', 'Machine Learning Algorithms', 'MLflow', 'Feature Engineering', 'Unsupervised Learning', 'Ethical AI', 'Artificial Intelligence Risk', 'Supervised Learning', 'Cudnn', 'Random Forest Algorithm', 'Genetic Algorithm', 'Speech Synthesis', 'Voice Interaction', 'Oracle Autonomous Database', 'Nvidia Jetson', 'Activity Recognition', 'Perceptron', 'Deeplearning4j', 'Operationalizing AI', 'Multi-Agent Systems', 'Voice Assistant Technology', 'Cognitive Automation', 'Knowledge-Based Systems', 'Speech Recognition Software', 'Hidden Markov Model', 'Microsoft Cognitive Toolkit (CNTK)', 'Kaldi', 'Artificial Intelligence Development', 'Expert Systems', 'Inference Engine', 'Intelligent Virtual Assistant', 'Objective Function', 'DALL-E Image Generator', 'Deep Learning', 'General-Purpose Computing On Graphics Processing Units', 'Azure Machine Learning', 'Variational Autoencoders', 'Association Rule Learning', 'Machine Learning Model Monitoring And Evaluation', 'Test Datasets', 'ChatGPT', 'K-Nearest Neighbors Algorithm', '3D Reconstruction', 'Loss Functions', 'LightGBM', 'Cortana', 'OmniPage', 'Machine Learning', 'Intelligent Automation', 'Recurrent Neural Network (RNN)', 'Artificial Neural Networks', 'Convolutional Neural Networks', 'Torch (Machine Learning)', 'Baidu', 'Support Vector Machine', 'Amazon Textract', 'Gradient Boosting', 'Collaborative Filtering', 'Embedded Intelligence', 'Automated Machine Learning', 'Apache MXNet', 'ModelOps', 'Ensemble Methods', 'Kernel Methods', 'Deck.gl', 'Microsoft LUIS', 'Confusion Matrix', 'Natural Language User Interface', 'Xgboost', 'Artificial Intelligence Systems', 'Reasoning Systems', 'Large Language Modeling', 'Sorting Algorithm', 'Interactive Kiosk', 'Soft Computing'

Reference: Lightcast Skills Categories, 2024 (https://lightcast.io/open-skills/categories)

Appendix 2 - List of skills used to identify green skills

Air quality and emissions

Air Permitting, Air Pollution Control, Air Quality, Air Quality Control, Air Sampling, Atmospheric Dispersion Modeling, Carbon Accounting, Carbon Footprint Reduction, Carbon Management, Carbon Monoxide Detectors, Continuous Emissions Monitoring Systems, Emission Calculations, Emission Reduction Projects, Emission Standards, Emission Testing, Emissions Controls, Emissions Inventory, Fugitive Emissions, Greenhouse Gas, Low Carbon Solutions, MACT Standards, National Emissions Standards For Hazardous Air Pollutants, Stack Emission Measurements, Vapour Recovery

Clean energy

Alternative Energy, Alternative Fuels, Biodiesel, Biodiesel Production, Biofuel Production, Biofuels, Biomass, Clean Technology, Geothermal Energy, Geothermal Heating, Methanol, Renewable Energy, Renewable Energy Development, Renewable Energy Markets, Renewable Energy Systems, Renewable Fuels

Climate change

Climate Analysis, Climate Change Adaptation, Climate Change Mitigation, Climate Change Programs, Climate Information, Climate Modeling, Climate Policy, Climate Resilience, Climate Variability And Change

Conservation

Conservation Biology, Conservation Planning, Environmental Impact Statements, Environmental Literacy, Environmental Protection, Environmental Risk Assessment, Environmentalism, Fish Conservation, Forest Conservation, Habitat Conservation, Habitat Conservation Plan, Low Impact Development, Marine Conservation, Rainwater Harvesting, Soil Conservation, Soil Genesis, Sustainability Planning, Threatened And Endangered Species Surveys, Water Conservation, Watershed Management, Wetland Conservation, Wetland Delineation, Wildlife Conservation, Wildlife Monitoring

Energy efficiency

Cooling Efficiency, Energy Analysis, Energy Conservation, Energy Conservation Measures, Energy Efficiency Analysis, Energy Efficiency Assessment, Energy Efficiency Improvement, Energy Efficiency Research, Energy Efficiency Services, Energy Efficiency Technologies, Energy Efficient Lighting, Energy Efficient Operations, Energy Modeling, Energy Saving Products, Energy-Efficient Buildings, Heat Recovery Steam Generators, Home Energy Assessment, LED Lamps,    Renewable Portfolio Standard, Residential Energy Conservation, Residential Energy Efficiency

Energy management

Advanced Distribution Automation, Automatic Meter Reading, Biomass Conversion, Biorefinery, Electric Meter, Electric Utility, Energy Analysis System, Energy Audits, Energy Consumption, Energy Conversion, Energy Demand Management, Energy Forecasting, Energy Management, Energy Management Planning, Energy Management Systems, Energy Market, Energy Policy, Energy Production, Energy Project Management, Energy Supply, Energy Transformation, Energy Transport, Flow Assurance, Fuel Metering, Gas Meter Systems, Hydraulic

Accumulators, Leadership in Energy and Environmental Design (LEED) Rating System, Load Shedding, Meter Reading, One-Line Diagram, Power System Simulator For Engineering, Public Utility, Resource Distribution, Smart Meter Installation, Smart Meter Systems, Sustainability Procedures, Transmission System Operator, Underground Utilities, Utility Cooperative

Environmental engineering and restoration

Biological Systems Engineering, Bioremediation, Ecological Engineering, Environmental Analysis, Environmental Contamination, Environmental Economics, Environmental Emergency, Environmental Field Services, Environmental Pollutants, Environmental Problem Solving, Environmental Program Management, Environmental Remediation, Environmental Technology, Environmental Toxicology, Geotextile, Land Reclamation, Landfill Design, Oil Containment Booms, Oil Skimmer, Oil Spill Contingency Plans, Pollution Control Systems, Reforestation, Remediation Systems, Restoration Ecology, Sanitary Engineering, Sediment Controls, Soil Contamination, Stream Restoration, Underground Storage Tanks (UST), Water Pollution, Wetland Restoration

Environmental regulations

Best Available Control Technology, California Environmental Quality Act (CEQA), Categorical Exclusions, Clean Water Act, Comprehensive Environmental Response Compensation and Liability Act (CERCLA), Emergency Planning And Community Right-To-Know Act, Endangered Species Act, Environmental Auditing, Environmental Compliance, Environmental Compliance Assessment, Environmental Due Diligence, Environmental Laws, Environmental Permitting, Environmental Protocols, EPA Regulations, Federal Insecticide Fungicide And Rodenticide Act, ISO 14000, ISO 14000 Series, ISO 14064, Marine Mammal Protection Act, Massachusetts Environmental Policy Act, National Environmental Policy Act, Natural Resources Law, Pollution Regulations, Resource Conservation And Recovery Act (RCRA), Restriction Of Hazardous Substances Directive, Safe Drinking Water Act, Spill Prevention Control And Countermeasure (SPCC), Total Maximum Daily Load, Water Law, Water Regulations Advisory Scheme

Nuclear energy

ANSI/ANS Standards, Monte Carlo N-Particle Transport Codes, Nuclear Core Design, Nuclear Criticality Safety, Nuclear Design, Nuclear Fuel, Nuclear Fuel Cycle, Nuclear Instrumentation Module, Nuclear Navy, Nuclear Plant Design, Nuclear Power, Nuclear Reactor, Nuclear Safety, Nuclear Technology, RELAP5-3D, Roentgen, Scintillator

Solar energy

Commercial Solar Projects, Concentrix Solar, Passive Solar Building Design, Photodetector, Photovoltaic Systems, Photovoltaics, PVsyst, Solar Application, Solar Cell Manufacturing, Solar Cells, Solar Consulting, Solar Design, Solar Development, Solar Energy, Solar Energy Systems Installation, Solar Engineering, Solar Equipment, Solar Inverter, Solar Manufacturing, Solar Panel Assembly, Solar Panels, Solar Photovoltaic Design, Solar Products, Solar Roofs, Solar Systems, Solar Thermal Installation, Solar Thermal Systems, Solar Water Heating

Waste management

E-Waste, Electrocoagulation, Landfill, Landfill Gas Collection, Leachate Management, Municipal Waste Management, Plastic Recycling, Recycling, Sludge, Sludge Disposal, Solid Waste Management, Tire Recycling,

Transfer Station, Trash Pickup, Waste Characterization, Waste Collection, Waste Disposal Systems, Waste Management, Waste Packaging, Waste Removal, Waste Sorting, Waste Tracking System, Waste Transport, Waste Treatment, Wastewater Treatment Plant Design

Water energy

Dam Construction, Hydraulic Structure, Hydroelectricity, Hydropower, WaterCAD

Wind energy

Wind Engineering, Wind Farm Construction, Wind Farm Design, Wind Farm Development, Wind Farming, Wind Power, Wind Turbine Maintenance, Wind Turbine Technology, Wind Turbines

Reference: Bruegel Twin Transition Dashboard, 2024

(https://www.bruegel.org/dataset/twin-transition-skills-dashboard)

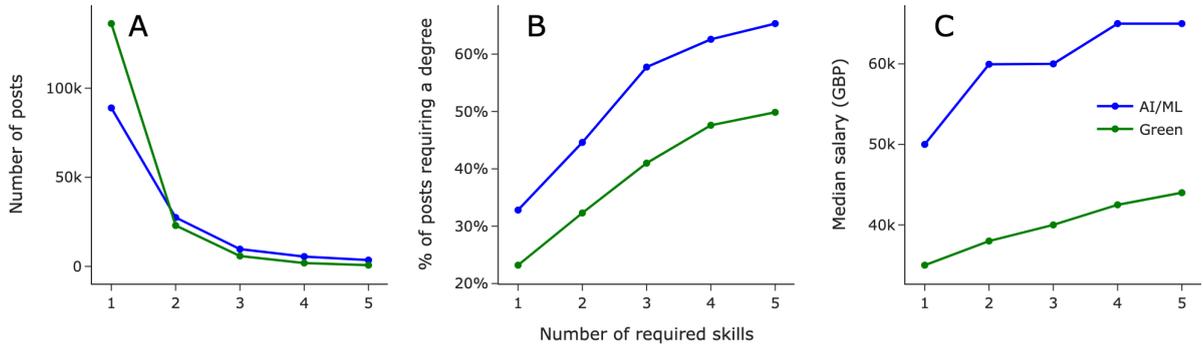

***Figure A1.*** *When do we define OJVs as related to AI or green? For our analysis, we label a posting as AI or green once it asks for at least one skill from the respective domain. The reason for this decision is that the number of labelled postings would decline quickly if one were to set a higher threshold of two or more skills, see panel A. On the other hand, we notice that for both AI and green postings, the educational requirements (B) and the offered wages (C) increases with the number of respective skills required in the OJVs.*

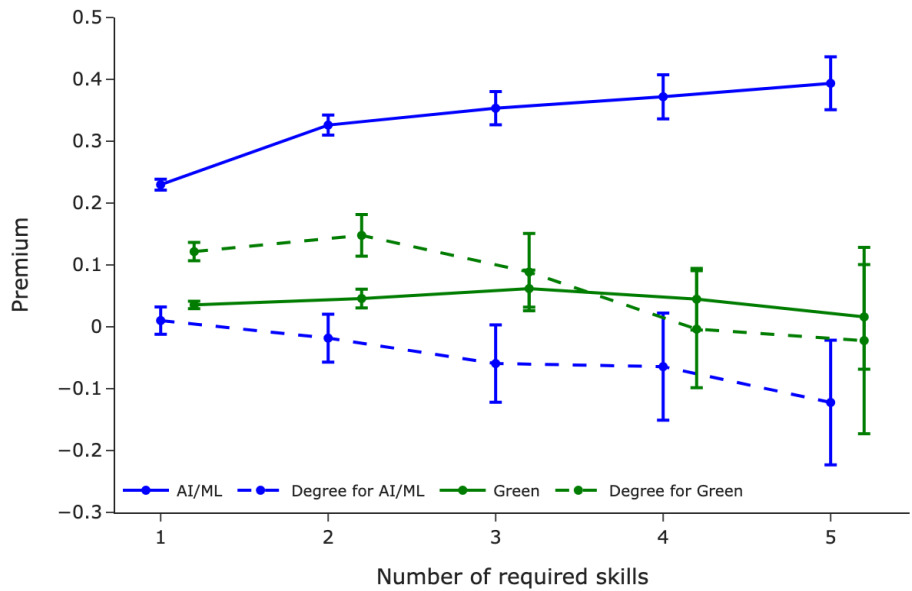

***Figure A2.*** *When we increase the number of required skills to define an AI or Green OJV, we see decreasing premiums for education (that is, requiring an HNC/HND or higher) and an increasing skill premium for AI roles. The premium for green roles stays relatively stable as the definition becomes more stringent.*

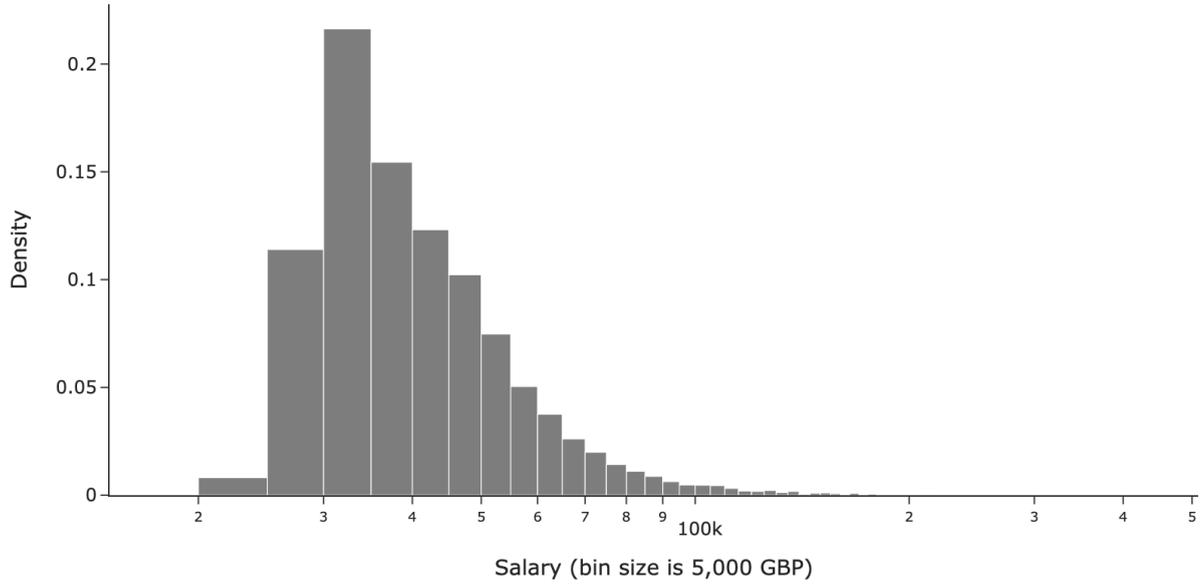

***Figure A3***. *Roughly 38.6 percent (4,187,603) of all job postings in our analysis contain wage information. Wages are log-normal distributed and slightly right-skewed. Therefore, we consider the log of wages for the subsequent analysis.*

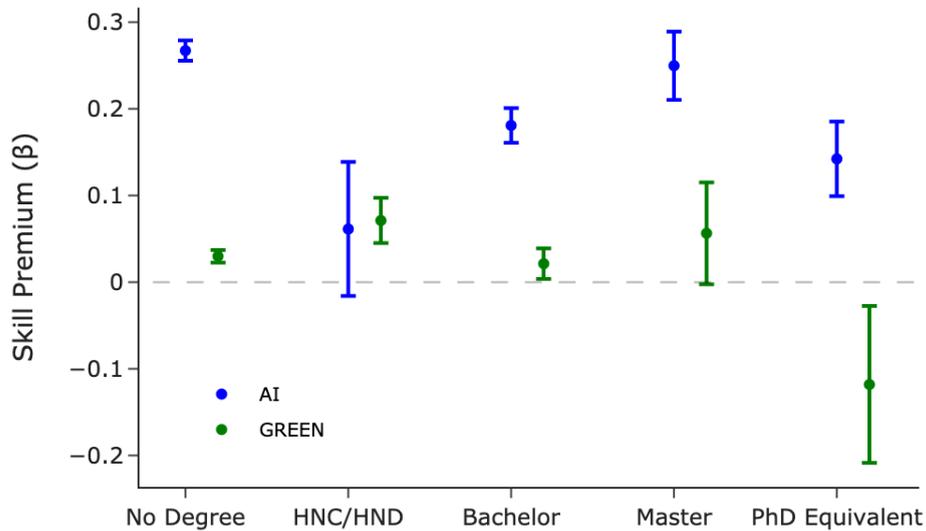

***Figure A4***. *This shows the AI/ML and Green premiums across degree requirements. It shows that AI premiums are consistently positive with the potential exception of HNC/HND-level postings. The premium for Green skills is less clear for Bachelor's and Master's degrees and becomes negative for PhDs. This could be due to many environmental researchers holding lower-paying academic positions.*

### Machine Learning Engineer

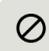 ★★★☆☆ 6 reviews

London

£48,237 a year - Fixed term contract

You must create an Indeed account before continuing to the company website to apply

**Apply on company site**

**Role Responsibility**

- Determining and refining machine learning objectives and run machine learning tests and experiments.
- Designing machine learning systems and self-running artificial intelligence software to automate predictive models.
- Solving complex problems with multi-layered data sets, as well as optimizing existing machine learning libraries and frameworks.
- Transforming data science prototypes and applying appropriate ML algorithms and tools.
- Proactively engaging with colleagues and maintain effective relationships and networks.

**The Ideal Candidate**

You will need demonstrable experience in and passion for machine learning and understanding of business user needs. You will have a Professional qualification or bachelor's degree in computer science, data science, mathematics, or a related field. You will have extensive knowledge of ML frameworks, libraries, data structures, data modelling, and software architecture. You should have an analytical mind and business acumen and be able to work competently and collaboratively as part of the Data & Insights Team and National Gambling Support Network providers and commissioners.

*Figure A5*. This is a job advertisement from the Greater London area. Information about the salary (highlighted in red), required education (blue), or skills (orange) had been extracted from all parts of the text of the posting, including headlines.

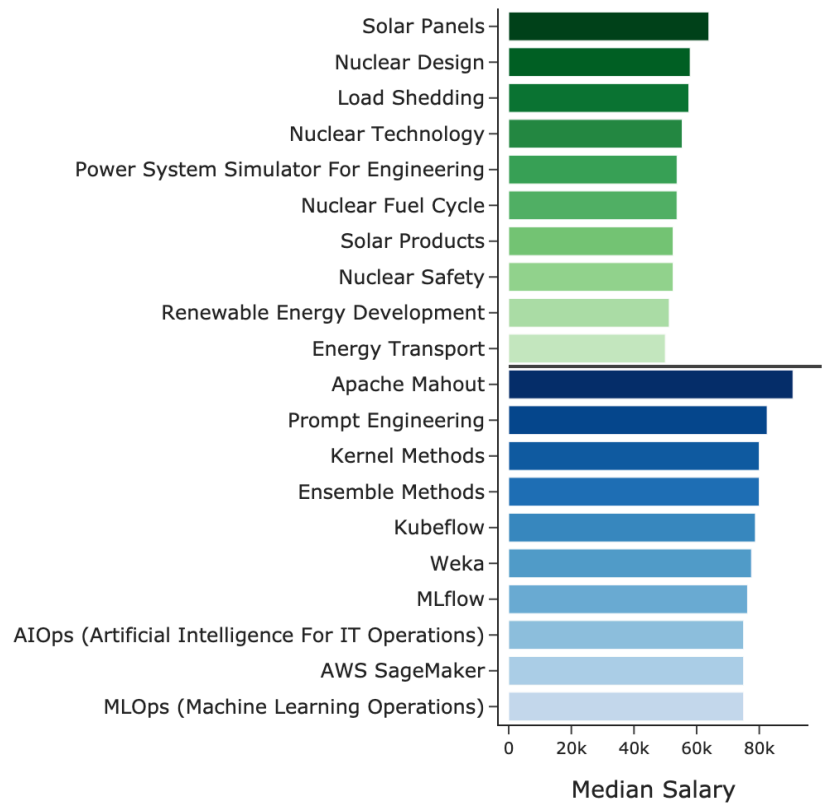

*Figure A5.* These are the skills with the highest paid associated salaries. It is the top median salary per skill after filtering for skills with at least 10 related posts having salary information.

**Table A1**. We compare the educational premium across degree levels for AI/ML, Green, and Other (non-AI, non-Green) postings. We find that there is a clear education gradient for Other job roles (model 6) which flattens for AI and Green roles (models 1 and 3). Interestingly, The premium for HNC/HND and Bachelor degrees becomes negative for AI roles, indicating that posts that don't require degrees are typically looking for individuals with more skills than these degrees typically offer. We also find that controlling for experience reveals an educational gradient for AI/ML roles, though the premiums for HNC/HND roles are still negative, indicating that experience is a dominant factor in determining wages for AI/ML roles.

Full Regression Results for Figure 2C1

| | AI/ML | | Green | | Other | |
|---|---|---|---|---|---|---|
| | (1) | (2) | (3) | (4) | (5) | (6) |
| Dep. Var. | | | *log(salary)* | | | |
| **Variables** | | | | | | |
| *hnc/hnd* | -0.14*** | -0.11** | 0.13*** | 0.14*** | 0.07*** | 0.08*** |
| | (0.03) | (0.05) | (0.01) | (0.01) | (0.00) | (0.00) |
| *bachelor* | -0.09*** | -0.02* | 0.09*** | 0.11*** | 0.07*** | 0.08*** |
| | (0.01) | (0.01) | (0.01) | (0.01) | (0.00) | (0.00) |
| *master* | 0.03*** | 0.09*** | 0.13*** | 0.21*** | 0.11*** | 0.12*** |
| | (0.01) | (0.02) | (0.01) | (0.03) | (0.00) | (0.00) |
| *phd equivalent* | -0.00 | 0.11*** | 0.12*** | 0.19*** | 0.29*** | 0.35*** |
| | (0.01) | (0.03) | (0.02) | (0.04) | (0.00) | (0.01) |
| *experience* | | 0.02*** | | 0.01*** | | 0.01*** |
| | | (0.00) | | (0.00) | | (0.00) |
| *intercept* | -30.05*** | -41.15*** | -62.37*** | -59.84*** | -45.30*** | -50.86*** |
| | (3.10) | (6.77) | (1.96) | (3.84) | (0.22) | (0.44) |
| **Fixed Effects** | | | | | | |
| *year* | yes | yes | yes | yes | yes | yes |
| *sic 1* | yes | yes | yes | yes | yes | yes |
| *soc 1* | yes | yes | yes | yes | yes | yes |
| *nuts 1* | yes | yes | yes | yes | yes | yes |
| R-squared | 0.35 | 0.34 | 0.29 | 0.29 | 0.34 | 0.34 |
| R-squared Adj. | 0.35 | 0.34 | 0.29 | 0.29 | 0.34 | 0.34 |
| N | 25286 | 5542 | 45063 | 11564 | 3819853 | 882126 |

Standard errors in parentheses, * $p < .1$, ** $p < .05$, *** $p < .01$

**Table A2**. Here we can see the change in AI, Green, and educational premiums from 2018-2023.

Full Regression Results for Figure 3A1

| Year | (2018) | (2019) | (2020) | (2021) | (2022) | (2023) |
|---|---|---|---|---|---|---|
| Dep. Var. | | | *log(salary)* | | | |
| **Variables** | | | | | | |
| *ai* | 0.21*** | 0.25*** | 0.24*** | 0.25*** | 0.26*** | 0.18*** |
| | (0.02) | (0.01) | (0.01) | (0.01) | (0.01) | (0.01) |
| *green* | 0.03*** | 0.03*** | 0.00 | 0.01 | 0.02** | 0.08*** |
| | (0.01) | (0.01) | (0.01) | (0.01) | (0.01) | (0.01) |
| *hnc/hnd* | 0.08*** | 0.11*** | 0.06*** | 0.07*** | 0.07*** | 0.09*** |
| | (0.01) | (0.01) | (0.01) | (0.01) | (0.01) | (0.00) |
| *bachelor* | 0.06*** | 0.07*** | 0.08*** | 0.07*** | 0.07*** | 0.10*** |
| | (0.00) | (0.00) | (0.00) | (0.00) | (0.00) | (0.00) |
| *master* | 0.10*** | 0.15*** | 0.17*** | 0.15*** | 0.05*** | 0.16*** |
| | (0.01) | (0.01) | (0.01) | (0.01) | (0.01) | (0.01) |
| *phd equivalent* | 0.34*** | 0.36*** | 0.34*** | 0.32*** | 0.23*** | 0.41*** |
| | (0.02) | (0.02) | (0.02) | (0.01) | (0.01) | (0.01) |
| *experience* | 0.01*** | 0.01*** | 0.01*** | 0.01*** | 0.01*** | 0.01*** |
| | (0.00) | (0.00) | (0.00) | (0.00) | (0.00) | (0.00) |
| *intercept* | 9.96*** | 10.01*** | 10.00*** | 10.00*** | 10.05*** | 10.08*** |
| | (0.01) | (0.01) | (0.01) | (0.01) | (0.00) | (0.00) |
| **Fixed Effects** | | | | | | |
| *sic 1* | yes | yes | yes | yes | yes | yes |
| *soc 1* | yes | yes | yes | yes | yes | yes |
| *nuts 1* | yes | yes | yes | yes | yes | yes |
| $r^2$ | 0.34 | 0.33 | 0.34 | 0.35 | 0.37 | 0.35 |
| $r^2$ Adj. | 0.34 | 0.33 | 0.34 | 0.35 | 0.37 | 0.35 |
| N | 120831 | 103601 | 108306 | 145661 | 188105 | 232728 |

Standard errors in parentheses, * $p < .1$, ** $p < .05$, *** $p < .01$

**Table A3**. We see that OJV data is mostly representative of the UK labour force regarding geography, industries, and occupations. However, there are some notable exceptions: The capital, London, is slightly overrepresented in our data. This is also the region with the largest share of AI and Green postings. For industries, business admin and support services are significantly more prominent in OJV in contrast to motor trades, wholesale, and retail, which is underrepresented. Lastly, OJVs show more postings in professional occupations and less than average job offers for elementary occupations. The most notable overrepresentation is the Administration & Support Services Industry which is a data artefact. A large proportion of job postings come from recruitment firms that are in this category.

| | % all posts[†] | | % of occ | | wage info |
|---|---|---|---|---|---|
| | ONS | OJV | AI | Green | Avg: 38.6 |
| **Regions (NUTS 1)** | | | | | |
| East Midlands | 7.3 | 6.2 | .4 | 1.1 | 54.8 |
| East of England | 9.6 | 8.8 | .7 | 1.1 | 52.5 |
| London | 14.4 | 20.3 | 1.7 | 1.0 | 48.1 |
| North East | 3.7 | 2.6 | .5 | 1.3 | 51.3 |
| North West | 10.9 | 10.7 | .5 | 1.2 | 53.1 |
| Northern Ireland | 2.6 | 1.8 | .8 | 1.4 | 47.8 |
| Scotland | 8.1 | 6.9 | .6 | 1.7 | 49.3 |
| South East | 14.1 | 14.9 | .6 | 1.2 | 52.7 |
| South West | 8.3 | 8.5 | .5 | 1.4 | 52.8 |
| Wales | 4.4 | 3.0 | 1.7 | 1.2 | 53.5 |
| West Midlands | 8.6 | 9.1 | .4 | 1.2 | 56.0 |
| Yorkshire and The Humber | 8.0 | 7.4 | .5 | 1.1 | 51.8 |
| NULL | | 25.0 | .8 | 1.3 | 0 |
| **Industries (SIC 1)** | | | | | |
| Agriculture, Forestry, and Fishing (A) | 1.7 | .2 | .8 | .9 | 20.8 |
| Mining, Quarrying, and Utilities (B,D,E) | 1.3 | .7 | 1.4 | 10.7 | 29.9 |
| Manufacturing (C) | 7.5 | 2.8 | 1.1 | 2.1 | 25.9 |
| Construction (F) | 5.0 | 1.7 | .4 | 2.6 | 29.9 |
| Motor Trades, Wholesale, Retail (G) | 14.4 | 4.8 | .5 | .7 | 25.4 |
| Transport & Storage (inc Postal) (H) | 5.0 | 1.4 | 1.1 | 1.2 | 32.8 |
| Accommodation & Food Services (I) | 7.4 | 3.1 | .2 | .7 | 27.3 |
| Information & Communication (J) | 4.3 | 4.2 | 2.8 | 1.0 | 25.2 |
| Finance & Insurance (K) | 3.4 | 2.0 | 2.2 | 1.0 | 23.2 |
| Real Estate Activities (L) | 1.9 | 1.2 | .6 | 1.3 | 35.2 |
| Professional, Scientific & Technical (M) | 8.8 | 6.7 | 1.3 | 2.2 | 25.6 |
| Administration & Support Services (N) | 8.7 | 52.2 | .6 | 1.2 | 43.9 |
| Education (P) | 8.5 | 3.1 | 2.5 | 1.2 | 48.9 |
| Health (Q) | 13.4 | 10.5 | .2 | .3 | 53.2 |
| Public Administration (O) | 4.5 | 2.7 | .8 | 1.7 | 53.4 |
| Other (R,S,T,U) | 4.3 | 1.6 | .8 | 1.4 | 35.3 |
| NULL | | 22.3 | .8 | 1.1 | 34.8 |
| **Occupations (SOC 1)** | | | | | |
| Managers, Directors, and Senior Officials | 10.8 | 10.1 | .6 | 1.3 | 38.4 |
| Professional Occupations | 20.1 | 33.6 | 1.3 | 1.3 | 38.1 |
| Associate Professional and Technical Occupations | 14.5 | 16.9 | 1.4 | 1.6 | 37.7 |
| Administrative and Secretarial Occupations | 10.4 | 8.8 | .2 | .9 | 43.1 |
| Skilled Trades Occupations | 10.1 | 6.4 | .2 | 1.6 | 40.0 |
| Caring, Leisure, and Other Service Occupations | 9.2 | 6.0 | .1 | .4 | 39.6 |
| Sales and Customer Service Occupations | 7.6 | 7.2 | .4 | .8 | 38.1 |
| Process, Plant and Machine Operatives | 6.4 | 4.4 | .2 | 1.4 | 42.7 |
| Elementary Occupations | 10.7 | 6.6 | .1 | 1.2 | 39.5 |
| NULL | | .3 | 1.7 | 1.0 | 0 |

[†]source: *Office of National Statistics*.
The non-null % are calculated after removing the NULL labeled posts. The most recent available data was from 2021 for regional and industry distributions and 2018 for the occupational distribution.

**Table A4.** We look at the relationship between the AI premium, Green premium, and degree premium across the SOC2 occupations and relate them to the demand of AI and Green skills as well as the percent of OJVs requiring a bachelor's degree or above.

| soc 2 occupation | ai $\beta$ | green $\beta$ | degree $\beta$ | % ai | % green | % degreed |
|---|---|---|---|---|---|---|
| Administrative | 0.20*** | 0.10*** | 0.20*** | 0.23 | 0.99 | 5.36 |
| Business and public service | 0.36*** | 0.16*** | 0.11*** | 1.77 | 1.84 | 12.95 |
| Business, media and public service | 0.16*** | 0.02 | 0.06*** | 0.90 | 1.25 | 19.43 |
| Caring personal service | -0.00 | 0.02 | 0.17*** | 0.12 | 0.22 | 3.22 |
| Community and civil enforcement | 0.00 | -0.13 | 0.46*** | 0.12 | 2.40 | 1.81 |
| Corporate managers and directors | 0.21*** | 0.07*** | 0.15*** | 0.69 | 1.24 | 12.34 |
| Culture, media and sports | 0.10 | 0.05 | 0.04*** | 0.55 | 0.65 | 12.06 |
| Customer service | -0.02 | -0.03 | 0.18*** | 0.22 | 0.59 | 2.79 |
| Elementary administration and service | 0.11*** | 0.03*** | 0.15*** | 0.13 | 1.24 | 1.97 |
| Elementary trades and related | 0.02 | 0.07* | 0.28*** | 0.20 | 0.78 | 1.41 |
| Health | 0.04 | -0.22*** | 0.02*** | 0.14 | 0.09 | 19.11 |
| Health and social care | 0.16*** | 0.04 | 0.32*** | 0.25 | 0.31 | 12.28 |
| Leisure, travel and related services | -0.08 | -0.03 | 0.26*** | 0.19 | 0.98 | 2.46 |
| Other managers and proprietors | 0.18*** | 0.01 | 0.26*** | 0.32 | 1.55 | 7.02 |
| Process, plant and machine | 0.20*** | 0.10*** | 0.20*** | 0.21 | 1.54 | 2.31 |
| Protective service | 0.17 | -0.03 | 0.20*** | 0.38 | 1.13 | 6.74 |
| Sales | 0.13*** | 0.10*** | 0.14*** | 0.55 | 0.83 | 5.49 |
| Science, engineering and tech | 0.36*** | 0.02** | 0.11*** | 1.25 | 1.87 | 8.34 |
| Science, research, engineering and tech | 0.18*** | -0.10*** | -0.02*** | 2.43 | 2.25 | 15.42 |
| Secretarial and related | 0.11*** | 0.06** | 0.17*** | 0.23 | 0.44 | 2.70 |
| Skilled agricultural and related trades | -0.16 | 0.00 | 0.37*** | 0.29 | 1.86 | 3.44 |
| Skilled construction and building trades | -0.10 | -0.04** | 0.23*** | 0.12 | 2.12 | 1.93 |
| Skilled metal, electrical and electronic | -0.09* | 0.09*** | 0.21*** | 0.28 | 1.78 | 2.09 |
| Teaching and other educational | 0.05** | 0.04 | 0.10*** | 0.47 | 0.29 | 20.45 |
| Textiles, printing and other trades | 0.06 | 0.01 | 0.07*** | 0.07 | 0.59 | 2.53 |
| Transport and mobile machine drivers | -0.08 | -0.05*** | 0.06*** | 0.11 | 1.22 | 1.19 |

This displays results of a regression ran across all soc2 occupations: $log(salary) \sim ai + green + degree + year + experience + sic\ 1 + nuts\ 1$ where the first three variables of interest are dummies indicating a post requires ai, green skills, or a bachelor's or above, respectively. This highlights that, while AI skills are often required more often in occupations with higher degree requirements, the premium for AI skills is often higher than the degree premium in these occupations like *Business and Public Service* and *Science, research, engineering and tech* while many occupations with high degree requirements rarely require AI skills like *Health* and *Teaching and other educational service*